\documentclass[12pt,english]{article}
\usepackage[english]{babel}
\usepackage{amsmath,amssymb,amscd,color}
\usepackage{amsfonts}
\usepackage{graphicx}
\usepackage{url}
\usepackage{hyperref}
\usepackage{cite}
\usepackage{physics}

\makeatletter
\renewcommand\section{\@startsection {section}{1}{\z@}%
                                   {-5.5ex \@plus -1ex \@minus -.2ex}%nn
                                   {2.3ex \@plus.2ex}%
                                   {\normalfont\large\bfseries}}
\renewcommand\subsection{\@startsection{subsection}{2}{\z@}%
                                     {-3.25ex\@plus -1ex \@minus -.2ex}%
                                     {1.5ex \@plus .2ex}%
                                     {\normalfont\bfseries}}

\numberwithin{equation}{section}

\makeatother

\newcommand{\bea}{\begin{eqnarray}}
\newcommand{\eea}{\end{eqnarray}}
\newcommand{\be}{\begin{equation}}
\newcommand{\ee}{\end{equation}}

\newcommand{\Z}{{\mathbb Z}}

\newcommand{\C}{{\mathbb C}}

\def\eg{{\it e.g.~}}

\newcommand{\cG}{{\cal G }}
\newcommand{\cN}{{\cal N }}

\newcommand{\cS}{{\cal S }}

\newcommand{\zb}{{\bar z}}

\newcommand{\hb}{{\bar h}}

\newcommand{\Jsf}{{\textrm J}}

\renewcommand{\title}[1]{\vbox{\center\LARGE{#1}}\vspace{5mm}}
\renewcommand{\author}[1]{\vbox{\center#1}\vspace{5mm}}
\newcommand{\address}[1]{\vbox{\center\footnotesize\em#1}}
\newcommand{\email}[1]{\vbox{\center\footnotesize\tt#1}\vspace{5mm}}

\begin{document}

\begin{titlepage}

 \begin{flushright}

\end{flushright}

\begin{center}

\hfill \\
\hfill \\
\vskip 1cm

\title{Conformal Perturbation Theory on K3:\\ The Quartic Gepner Point}

\author{Christoph A. Keller
}

\address{
Department of Mathematics, University of Arizona, Tuscon, AZ 85721-0089, USA}

\email{ cakeller@math.arizona.edu}

\end{center}

\vfill

\abstract{The Gepner model $(2)^4$ describes the sigma model of the Fermat quartic K3 surface. Moving through the nearby moduli space using conformal perturbation theory,  we investigate how the conformal weights of its fields change at first and second order and find approximate minima. This serves as a toy model for a mechanism that could produce new chiral fields and possibly new nearby rational CFTs.}

% \abstract{The Gepner model $(2)^4$ describes the sigma model of the Fermat quartic K3 surface. We explore its nearby moduli space using conformal perturbation theory. Our results illustrate the interplay between first and second order contributions and allow us to study approximate nearby minima of the conformal weights of fields. This toy model illustrates a mechanism that could produce new chiral fields and possibly new nearby rational CFTs.}

\vfill

\end{titlepage}

\eject

\tableofcontents
\newpage

\section{Introduction}
Conformal perturbation theory has a long history \cite{Kadanoff:1978pv, KADANOFF197939,Zamolodchikov:1987ti}. One of its uses is to explore the conformal manifold of a CFT \cite{Kutasov:1988xb,Friedan:2012hi}.
How to do this is well understood in principle, but in practice performing concrete computations can be quite challenging. The first challenge is to find a  CFT that has an actual conformal manifold, that is whose spectrum contains exactly marginal operators. Such theories are not that common; in fact it is believed that the only examples are theories with a global $U(1)$ symmetry, or theories with enough supersymmetry, namely $N=2$ supersymmetry \cite{Dixon:1987bg,Lerche:1989uy,Gukov:2004ym}. 

The second challenge is to have explicit expressions for the correlation functions. At first order perturbation, this is not an issue since the form of the 3pt functions is fixed by conformal symmetry. The perturbation integral can thus be evaluated directly and the result depends only on a single parameter of the theory, namely the 3pt constant \cite{Dijkgraaf:1987jt,Cardy:1989da}. For higher order perturbations however, conformal invariance no longer fixes the functional form of the higher point correlation functions. Their precise form thus depends on the dynamics of the CFT.
The third challenge is then to integrate these correlation functions. This poses several technical problems. First it is necessary to introduce a regularization scheme to deal with divergences. The choice of this scheme does not matter much at first order, but becomes important at higher order, as it is related to the choice of coordinates on the conformal manifold. Then an efficient way of evaluating the resulting integrals needs to be found.

In practice, these challenges constrain the types of examples we can work with. Because of point one, we need to find a CFT with $N=2$ supersymmetry. It is of course believed that Calabi-Yau sigma models provide a vast class of such examples. However, because of point two, these are mostly unusable for our purposes because we do not have explicit expressions for the their correlators. There are effectively only two types of models whose correlators are known well enough so that we can work with them: torus theories and their orbifolds, and rational theories.

Perturbations of torus orbifolds were considered in \cite{Eberle:2001jq,Keller:2019suk,Keller:2019yrr,Benjamin:2020flm}.  Deforming by a modulus in the untwisted sector is not very interesting, as it simply deforms the underlying torus, giving another torus orbifold. To have interesting perturbations, it is thus necessary to deform by a modulus in the twisted sector.  The twist sector selection rules however then imply that all first order contributions vanish. It is thus necessary to focus on the second order contributions, which are indeed non-vanishing.

In this article however we want to investigate examples with  first and second order contributions that are both non-vanishing. For the reasons given above, this rules out torus orbifolds.
We therefore focus on CY sigma models described by rational CFTs that are not torus orbifolds. Luckily, there are such theories: the so-called Gepner models \cite{Gepner:1987qi,Gepner:1989gr}. Such models are based on $N=2$ minimal models. These minimal models have central charge $c= 3k/(k+2)$, and they are rational with respect to the $N=2$ superconformal algebra, which in particular means that all their dynamical data such as correlation functions can be computed explicitly. 
Gepner models that describe the sigma model of a CY $D$-fold are then constructed as the tensor product of such minimal models with total central charge 3$D$ whose symmetry algebra is extended by the spectral flow operator.  This (worldsheet) spectral flow operator corresponds to the spacetime supersymmetry operator, and in particular imposes integrality of the $U(1)$ charges. Because of this, Gepner models give supersymmetric string theories. In fact, \cite{Gepner:1987qi,Gepner:1989gr} argued that they correspond to specific points in the moduli space of Calabi-Yau compactifications, namely to Fermat curves. Such Gepner models thus provide theories that do have a conformal manifold and correlation functions that are known explicitly, which makes them suitable for the purposes of this article.

Our goal in this article is twofold. First, we want to establish the framework for conformal perturbation theory of Gepner models and use it to explore their moduli space neighborhood. For this we need to compute correlation functions and evaluate the perturbation integrals.
Our second goal is motivated by a more specific question: Are rational CFTs dense in this moduli space? The motivation for this question comes from torus compactifications, where the answer is yes \cite{Moore:1998pn,Wendland:2000ye,Hosono:2002yb}.
\cite{Gukov:2002nw} then conjectured that rational CFTs are dense also in the moduli space if the CY is a K3, but not if it is a higher dimensional CY manifold. This conjecture was approached from a geometric perspective in \cite{Chen:2005gm,Kidambi:2022wvh,Okada:2022jnq}. Here we want to approach it using conformal perturbation theory.\footnote{A similar question was recently studied in \cite{Antunes:2022vtb} using perturbative RG flows.} 

From the point of view of perturbation theory, how could a new rational CFT arise near a Gepner point? Because $c=6$ or higher, this CFT cannot be rational with respect to the overall $N=(2,2)$ algebra. This means that there must be additional chiral fields in the CFT, that is fields with $\bar h=0$. At the Gepner point itself, such additional fields were of course given by the $N=(2,2)$ algebra of the individual factors; however, we do not expect these to survive the perturbation.
The idea is therefore to start with a field that is not chiral at the Gepner point with say weight $(h(0),\bar h(0))$. Under perturbation by a modulus with coupling constant $\lambda$, we are hoping to find a nearby point $\lambda_0$ such that 
\be
\bar h(\lambda_0)= 0\ .
\ee
The symmetry algebra at $\lambda_0$ is thus enhanced by this new chiral field. If we are lucky, then the CFT at $\lambda_0$ might indeed be rational. Even though not sufficient, having such a field is definitely necessary for this. See for instance \cite{Benjamin:2020flm} for a description of this mechanism for the toy model of a free boson on $S^1$.

In this article we investigate a toy version of this mechanism. Note that since unitarity imposes $\bar h \geq 0$, it is clear that $\bar h(\lambda)$ has to be a minimum at $\lambda_0$.  We will try to identify such minima for certain fields, regardless of whether $\bar h=0$ at the minimum. To this end we approximate $\bar h(\lambda)$ by a quadratic function. For it to have an interesting minimum, we need both the linear and the quadratic term of that function to be non-vanishing. That is, we want a contribution both from first and second order perturbation theory.

To achieve such a situation, we consider the Gepner model $(2)^4$; that is, the $N=(2,2)$ CFT with central charge $c=6$ consisting of four copies of the $k=2$ minimal model. Geometrically, this corresponds to the sigma model of the K3 given by the quartic Fermat surface
\be
X^4+Y^4+Z^4+W^4=0\ .
\ee
In a sense this is the simplest Gepner model. Its correlators are easier to compute due to the fact that it consists of copies of the $k=2$ minimal model only. This minimal model has $c=3/2$, and is in fact equivalent to the Ising model tensored with a free boson. Its correlation functions are therefore well known.

We investigate the lifting of the lightest non-BPS states in the spectrum of this  Gepner point. It turns out that there are 12 such states with $h=\bar h =1/4$. 
The moduli space of K3 is given by $O(4)\times O(20)\backslash O(4,20)/O(4,20;\mathbb{Z})$ (see for instance \cite{Aspinwall:1996mn} for a review). There are thus 80 real moduli, or, in $\cN=4$ language, 20 $\cN=4$ multiplets of moduli. It is straightforward to identify them in our Gepner model, and it turns out that they naturally fall into three types which we call A,B and C.

We first investigate the lifting of the 12 lightest states at first order. For this, we pick the $(c,c)$ moduli for each of the 20 $\cN=4$ multiplets. It turns out that the 2 type A moduli do not lift any of the states at first order, and the remaining 18 moduli each lift only 2 of the 12 states. The vanishing of so many correlation functions is not too surprising, since the Gepner point is a point of much enhanced symmetry; in particular there are four separate $U(1)$ charges that need to be preserved. But importantly, unlike for the torus orbifold case, there are states that are lifted at first order. Since we are interested in finding interesting minima, we focus only on these states.

Next we proceed to second order perturbation theory and compute the pertinent 4pt functions. Surprisingly, we find that the 6 type B moduli in our hard sphere  regularization scheme do not give a second order contribution to the lifting!
It is not clear to us what to make of this observation, as there does not seem to be a deep underlying reason for this. It is possible that this is an artefact of the specific Gepner model we investigate, or even just of the specific states we consider.

\begin{figure}[h!]
	\centering\includegraphics[width=.6\textwidth]{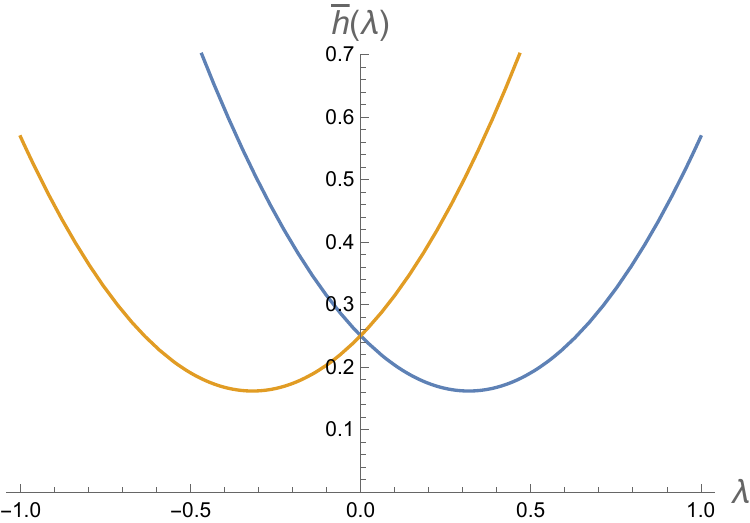}
	\caption{$\bar h$ as a function of the coupling constant $\lambda$ for the two lifted fields. Note the minima at $\lambda=\pm0.32$. }
	\label{fig:plotintro}
\end{figure}

More importantly however, the remaining 12 moduli of type C give a positive second order contribution. Each such modulus thus gives a second order approximation of $\hb(\lambda)$ for two states that indeed has a minimum near the Gepner point. These  minima turn out to be about $\hb=0.16$, which is significantly smaller than the starting value of $\hb =0.25$. These states thus provide a toy version of the symmetry enhancing mechanism described above.

What does all this tell us about rational CFTs in CY moduli spaces? We first remark that our computation is only a rough toy model for the symmetry enhancing mechanism. Since $\hb\neq 0$, the field does not lead to an enhanced symmetry at the point. Such an enhancement is anyway impossible for the scalar fields we are considering: they could at best become a second vacuum with $h=\hb=0$, which would of course violate cluster decomposition. Moreover our second order version of $\hb(\lambda)$ may not be a good approximation to the full function $\bar h(\lambda)$, and our minimum not a good approximation to the actual minimum. To test the denseness conjecture of \cite{Gukov:2002nw},  we have to perform our analysis with a field of non-zero spin. We would then want to repeat it for a CY 3-fold and compare to the K3 results. If we find a qualitative difference between the two, this could then be taken as evidence in favor of the conjecture. We will leave this to future work. As it is, our article illustrates how the symmetry enhancement mechanism can work in principle and how to do the necessary computations in practice.

\section{Conformal perturbation theory}
\subsection{Shift of conformal dimension}
Let us give a quick overview of conformal perturbation theory  \cite{Cardy:1987vr,Dijkgraaf:1987jt,Kutasov:1988xb}, following the exposition in \cite{Benjamin:2020flm}. We are interested in the shift of the conformal weight of primary fields under perturbation by an exactly marginal field $\Phi$, the modulus.
For this we schematically expand the two point function 
\be\label{2ptpert}
\langle\varphi^\dagger (z_1)\varphi(z_2)\rangle_\lambda = \langle\varphi^\dagger(z_1)\varphi(z_2) e^{\lambda\int d^2w \Phi(w)}\rangle
= \frac{1}{(z_1-z_2)^{2h(\lambda)}(\zb_1-\zb_2)^{2\hb(\lambda)}}
\ee
in powers of the coupling $\lambda$.  From the correlator (\ref{2ptpert}) we can read off the shift in the conformal dimensions $(h,\hb)$ of $\varphi$.  To be more precise, we are looking to compute the coefficients $h^{(n)}$ in the expansion
\be\label{hexpand}
h(\lambda)= \sum_{n=0}^\infty h^{(n)} \lambda^n\ ,
\ee
where $h^{(0)}$ is the dimension of the field in the unperturbed theory. Since the spin is integral and therefore constant under perturbations, the exact same expression (except for the $\hb^{(0)}$ term) must hold for $\hb(\lambda)$. 

Our discussion so far has been very schematic. In particular, we did not deal with the issue of divergences and their regularization. A more precise form of (\ref{2ptpert}) is
\be\label{2ptpertreg}
\langle\varphi^\dagger (z_1)\varphi(z_2)\rangle_\lambda = \langle\varphi^\dagger(z_1)\varphi(z_2) e^{\lambda\int_\varepsilon d^2w \Phi(w) + c.t.}\rangle
= \frac{A(\varepsilon, \lambda)}{(z_1-z_2)^{2h(\lambda)}(\zb_1-\zb_2)^{2\hb(\lambda)}}\ .
\ee
Here $\int_\varepsilon$ indicates a regularized version of the integral with regularization parameter $\varepsilon$ and $c.t.$ indicates some ($\varepsilon$ dependent) counterterms. $A(\varepsilon,\lambda)$ finally is the wave function renormalization of $\varphi$.

To find the coefficients in (\ref{hexpand}), we first write the right-hand side of (\ref{2ptpert}) as
\be\label{shiftexp}
\frac{1}{(z_1-z_2)^{2h^{(0)}}} \frac{1}{(\zb_1-\zb_2)^{2\hb^{(0)}}} \prod_{n=1}^\infty \exp(-2h^{(n)} \lambda^n\log|z_1-z_2|^2 )
\exp(-2a_n(\varepsilon) \lambda^n)
\ ,
\ee
where the $a_n(\varepsilon)$ are the appropriate expansion coefficients of $A(\varepsilon,\lambda)$ and we have used the fact that $h^{(n)}=\hb^{(n)}$ for $n\geq 1$. In fact, since $\varepsilon$ is the only dimensionful quantity, the $\varepsilon$ dependent part is necessarily of the form $a_n(\varepsilon) = - h^{(n)}\log \varepsilon^2 $, so that we can combine the terms to $-2h^{(n)}\lambda^n\log(|z_{12}|^2/\varepsilon^2)$.
Expanding the $z_{12}$ dependent part of this to second order in $\lambda$, we find
\be\label{hexpand2}
-2\log |z_{12}|^2 h^{(1)} \lambda +2(\log |z_{12}|^2 h^{(1)})^2\lambda^2 - 2\log |z_{12}|^2 h^{(2)}\lambda^2 +\ldots\ .
\ee

To fix the coefficients $h^{(n)}$, we compute the $n$-th term in the expansion of the left-hand side of (\ref{2ptpert}),
\be\label{ordern}
\frac{\lambda^n}{n!}\int d^2w_1 \ldots d^2 w_n \langle \varphi^\dagger(z_1)\varphi(z_2)\Phi(w_1)\ldots \Phi(w_n)\rangle\ .
\ee
A shift in $h$ thus occurs if the integral (\ref{ordern}) produces logarithmic terms $\log|z_1-z_2|$. 
This integral is divergent due to the singularities that arise when a modulus $\Phi$ approaches another field. It therefore needs to be regularized in some way. As usual, regularization will introduce a length scale and thus break manifest conformal invariance. Only if $\Phi$ is indeed exactly marginal will the perturbed theory still be conformal. In practice, this breaking shows up as the appearance of logarithmic terms $\log|z_1-z_2|$. These are of course exactly the type of terms that are needed to shift the conformal weight $h$.

The regularization scheme that we use is hard sphere regularization with minimal subtraction. This means we cut out little discs of radius $\varepsilon$ around all fields, and subtract any divergent powers of $\varepsilon$, leaving any constant terms untouched. This is a local regularization scheme since the discs cut out only depend on the insertion point of the colliding fields, and not on the positions $z_1,\ldots z_n$ of any other fields.
Note that we are primarily interested in logarithmic divergences, that is terms of the form $\log\varepsilon$. The reason for this is that because of the comment above, such terms will always be accompanied by logarithmic terms $\log|z_1-z_2|$. In practice it is thus a convenient shortcut to simply identify terms of the form $\log \varepsilon$. We will however briefly discuss how $\log|z_1-z_2|$ terms arise below. 

Let us briefly discuss the counterterms that appear in (\ref{2ptpertreg}). In general they are integrals of local fields with coefficients in $\varepsilon$ and $\lambda$. Their purpose is to cancel the divergences that appear when the moduli $\Phi$ collide with each other. Consider the order $\lambda^2$ counterterm. Let 
\be
\Phi(w_1)\Phi(w_2) \sim \sum_{\Delta_k\leq 2} \frac{C_{\Phi\Phi\Psi_k}}{|w_1-w_2|^{4-\Delta_k}} \Psi_k(w_2)
\ee
be the scalar singular part of the OPE of $\Phi$ with itself. In the integral (\ref{ordern}), the collision of $\Phi(w_1)$ with $\Phi(w_i)$ gives a divergence
\be\label{ctdivergence}
\int_{|w_1-w_i|>\varepsilon} d^2 w_1\langle \ldots \Phi(w_1)\Phi(w_i) \ldots \rangle = \sum_{\Delta_k\leq 2} \frac{2\pi C_{\Phi\Phi\Psi_k}}{2-\Delta_k} \varepsilon^{\Delta_k-2} \langle \ldots  \Psi_k(w_i)\ldots \rangle\ .
\ee
We want to use a minimal subtraction scheme for our counterterms. That is, we want to only cancel terms divergent in $\varepsilon$. This means that to order $\lambda^2$, we choose our counterterms as
\be
-\sum_{\Delta_k<2} \lambda^2 \frac{2\pi C_{\Phi\Phi\Psi_k}}{2-\Delta_k} \varepsilon^{\Delta_k-2} \int d^2 w \Psi_k(w)
+ \sum_{\Delta_k=2} \lambda^2 2\pi C_{\Phi\Phi\Psi_k} \log (\varepsilon)  \int d^2 w \Psi_k(w)\ .
\ee
These then cancel the divergences arising from (\ref{ctdivergence}) at second order in any perturbed correlator.

\subsection{Perturbation theory up to second order}\label{ss:2ndpert}
Let us now work out explicit expressions for perturbation theory up to second order.
At first order, eq. (\ref{ordern}) is of the form
\be
\lambda \int d^2w_1 \langle \varphi^\dagger(z_1)\varphi(z_2)\Phi(w_1)\rangle
= \lambda 2\pi C_{\varphi^\dagger\varphi\Phi} \log\left(\frac{|z_{12}|^2}{\varepsilon^2}\right) z_{12}^{-2h_\varphi}\zb_{12}^{-2\bar h_\varphi}\ .
\ee
The $\varepsilon$ dependence gives the wave function renormalization. 
Re-inserting the appropriate $z_{12}$ and comparing to (\ref{shiftexp}), we recover the well known result 
\be
h^{(1)} = -\pi C_{\varphi^\dagger\varphi\Phi}\ .
\ee
If there are multiple fields $\varphi_i$ with the same conformal dimension, then we have to perform degenerate perturbation theory to take into account operator mixing: $h^{(1)}$ becomes now the matrix
\be
H^{(1)}_{ij} = -\pi C_{\varphi_i^\dagger\varphi_j\Phi}\ ,
\ee
where $(h_{\varphi_i},\hb_{\varphi_i})=(h_{\varphi_j},\hb_{\varphi_j})$. We can then simply choose a new orthonormal basis for the fields $\varphi_i$ such that $H^{(i)}$ becomes diagonal.

To compute the second order shift, we perform a M\"obius transformation
\be\label{mobius}
f(z):=\frac{(z-z_2)(w_1-z_1)}{(z-z_1)(w_1-z_2)}\ ,
\ee
which sends $w_2$ to the cross ratio $x:=f(w_2)$. The second order integral then reads
\be
\frac{\lambda^2}{2}\int d^2 w_1\,z_{12}^{-2h_\varphi}\zb_{12}^{-2\bar h_\varphi}\left|\frac{z_1-z_2}{(z_1-w_1) (w_1-z_2)} \right|^2
\int d^2 x \,\langle \varphi^\dagger(\infty)\Phi(1)\Phi(x)\varphi(0)\rangle'\ .
\ee
Here the prime indicates an integral that is regularized using the hard sphere scheme described above. However, as pointed out in \cite{Keller:2019yrr},  we need to be careful about how this  regularization works for the integral after the conformal transformation (\ref{mobius}). The issue is that now the $x$ integral depends on $w_1$ and also on $z_1$ and $z_2$ due to the regulator, namely the small disk cut out at the insertion point $x$. After performing the conformal map (\ref{mobius}), the radius of the disks is
\be
\frac{w_1-z_1}{w_1-z_2} \frac{\varepsilon e^{i\theta}}{z_2-z_1} + \ldots\ .
\ee
Reference \cite{Keller:2019yrr} describes how to deal with this $w_1$ dependence. The idea is to introduce an auxiliary regularization scheme by subtracting the non-integrable divergences $G_{reg}$ of the correlator,
\be
G^{(r)}(x) := \langle \varphi^\dagger(\infty)\Phi(1)\Phi(x)\varphi(0)\rangle' - G_{reg}(x)\ .
\ee
Here $G_{reg}(x)$ is the sum of the contribution of fields that give a non-integrable divergence at $0,1,\infty$.
Up to suppressed powers of $\varepsilon$, the integral in this case is simply given by --- see \cite[section 2.3]{Keller:2019yrr}:
\begin{align}\label{2ndorderint2}
\frac{\lambda^2}{2}\int d^2 w_1\,z_{12}^{-2h_\varphi}\zb_{12}^{-2\bar h_\varphi}\left|\frac{z_1-z_2}{(z_1-w_1) (w_1-z_2)} \right|^2
\int d^2 x \,\langle \varphi^\dagger(\infty)\Phi(1)\Phi(x)\varphi(0)\rangle'\ \nonumber\\
= \pi\lambda^2 \log\left(\frac{|z_{12}|^2}{\varepsilon^2}\right) z_{12}^{-2h_\varphi}\zb_{12}^{-2\bar h_\varphi}
\int d^2 x G^{(r)}(x)\ .
\end{align}
This is then the `proper' contribution to the perturbation integral at second order. Reference \cite{Keller:2019yrr} argued that if power divergences in $\varepsilon$ appear, they will be removed by regularizing the integral and hence will not contribute to $h^{(2)}$.

In fact the argument \cite{Keller:2019yrr} needs to be slightly modified, as it assumed that there were no first order contributions. This is equivalent to saying that $G_{reg}(x)$ did not contain any logarithmic divergences. If there are such divergences, they will in fact lead to the $(\log z_{12})^2$ term in (\ref{hexpand2}).
To see this, insert an orthonormal basis of states into (\ref{ordern}) to obtain
\be
\frac{\lambda^2}{2!}\sum_{\phi}\int d^2w_1 d^2 w_2 \langle \varphi|\Phi(w_1)|\phi\rangle\langle\phi|\Phi(w_2)|\varphi\rangle\ .
\ee
In this expansion, the states $\phi$ that have the same dimensions as $\varphi$ will lead to a product of two logarithmic terms. Since we chose $\phi$ to diagonalize the first order lifting matrix, the contribution is of the form
\be
\lambda^2 2\pi^2 C^2_{\varphi^\dagger\varphi\Phi}\left(\log \varepsilon^{-2}\right)^2
= 2\lambda^2 (h^{(1)}\log \varepsilon^{-2})^2\ .
\ee
Once we reintroduce the $z_1$ and $z_2$ dependence, we immediately find that this is exactly the quadratic term of the expansion of the exponential of the first order shift in (\ref{shiftexp}). To get the linear term $h^{(2)}$ in the expansion of the exponential of the second order shift, we can subtract the contribution of the external fields $\varphi$ in the internal channel from the correlation function. In fact, \cite{Benjamin:2020flm} argued that this subtraction also works at higher order and is the analogue of the connected diagrams in standard QFT perturbation theory.

In total, we thus obtain
\be\label{hshift}
h(\lambda)= h^{(0)} -\pi C_{\varphi^\dagger\varphi\Phi}\lambda - \frac{\pi}2 M \lambda^2 + O(\lambda^3)\ ,
\ee
where 
\be\label{H0}
M = \int_{\C} d^2 x\,G^{(r)}(x)\ 
\ee
and $G^{(r)}$ is the integral of the suitably regulated four-point function in eq. (\ref{2ndorderint2}), or more precisely, its constant part \cite{Keller:2019suk}.

\section{$N=2$ minimal models}
\subsection{Primaries and fusion rules}
Let us now introduce the ingredients of the CFT that we will consider.
$N=2$ minimal models are CFTs that are rational with respect to the $N=2$ superconformal algebra. They have central charge
\be
c = \frac{3k}{k+2}
\ee
and can be constructed as cosets or as a tensor product of $\Z_k$ parafermions with a free boson \cite{Fateev:1985mm,Zamolodchikov:1986gh}. We will focus on the case $k=2$, where the parafermions become ordinary fermions, which in turn can be described by the Ising model. In the following we mostly follow the conventions of \cite{Stanishkov:2017rrr}, which provides a nice overview of $N=2$ minimal models.

In the NS sector, the primary fields are given by $N^l_m$, $l=0,1,\ldots k$ and $m=-l,-l+2,\ldots ,l$, with conformal weight and charge
\be
\Delta^l_m=\frac{l(l+2)-m^2}{4(k+2)}\ , \qquad Q^l_m = \frac{m}{k+2}\ .
\ee
In the R sector, the primary fields are given by $R^l_{m,\alpha}$, $l=0,1,\ldots k$, $m=-l,-l+2,\ldots ,l$ and $\alpha=\pm1$, with conformal weight and charge \footnote{Note our convention for the $U(1)$ charge differs by a factor of $\frac12$ from \cite{Stanishkov:2017rrr}. 
}
\be
\Delta^l_m=\frac{l(l+2)-(m+\alpha)^2}{4(k+2)}+\frac18\ , \qquad Q^l_m = \frac{m+\alpha}{k+2} -\alpha \frac12\ .
\ee
In what follows, we will focus on the NS sector. The fusion rules are
\be
[N^{l_1}_{m_1}]\times[N^{l_2}_{m_2}] = \sum_{l=|l_1-l_2|\textrm{\ step\ }2}^{\min(l_1+l_2,2k-l_1-l_2)} [\Psi^{l}_{m}]\ ,
\ee
where 
\bea
\Psi^{l}_{m}= (N^{l}_{m_1+m_2})^{even} && |m_1+m_2|\leq l\\
\Psi^{l}_{m}= (N^{k-l}_{m_1+m_2\pm(k+2)})^{odd} && |m_1+m_2|> l\ .
\eea
The appearance of two types of fusion channels, namely `odd' and `even' comes from the fact that the Ward identities do not allow to eliminate all $G$-descendants from three point functions of descendants: we will be left with either 0 or 1 $G$-descendant in the three point function. This means that the even fusion rules describe correlators that have an even total number of $G$-descendants, and odd fusions rules correlators with an odd total number.
Indeed we can check this on the level of $U(1)$ charges: We have
\be
Q(N^{l_1}_{m_1})+Q(N^{l_2}_{m_2})=Q(\Psi^{l}_{m})
\ee
in the even case, and 
\be
Q(N^{l_1}_{m_1})+Q(N^{l_2}_{m_2})=Q(\Psi^{l}_{m})\pm 1
\ee
in the odd case.

For future use, let us write down some specific fusion rules in the case of $k=2$:
\bea\label{fusion1}
N^1_1 \times N^1_1 &=& (N^2_2)^{even} + (N^2_{-2})^{odd}\\
N^1_1 \times N^1_{-1} &=& (N^0_0)^{even} + (N^2_{0})^{even}\\
N^2_2 \times N^1_1 &=& (N^1_{-1})^{odd}\\
N^2_2 \times N^1_{-1} &=& (N^1_{1})^{even}\\
N^2_2 \times N^2_2 &=& (N^2_0)^{odd}\label{f22}\\
N^2_2 \times N^2_{-2} &=& (N^0_0)^{even}\label{f2m2} \label{fusionlast}
\eea

\subsection{The bosonic subalgebra and characters}
In view of the Gepner construction below, it is useful to decompose the irreducible representations of the $N=2$ SCA into irreducible representations of its bosonic subalgebra, that is the algebra generated by its bosonic generators and even numbers of fermionic generators. The primaries above then split into two irreducible representations with even and odd fermion number respectively. This will be very useful once we introduce the GSO projection, for which we need to keep track of the fermion parity.
Moreover, this also allows to treat Ramond and NS representations on equal footing, since they both have trivial monodromy with respect to the bosonic subalgebra. 

For labelling purposes we therefore introduce an additional parameter $s=0,1,2,3$ that keeps track of the fermion parity and the Ramond and NS sectors. Here $s=0,2$ labels the two parities in the NS sector, and $s=1,3$ in the R sector. The irreps of the bosonic subalgebra are thus labelled by triples $(l,m,s)$ with identifications $m\sim m\pm (2k+4)$ and $s\sim s\pm4$. Moreover we have the field identification
\be\label{fieldid}
(l,m,s) \sim (k-l,m+k+2,s+2)\ .
\ee
It can therefore be useful to have the labels run as
\be\label{labelcond}
0\leq l \leq k, \qquad m=-k-1,\ldots,k+2 \qquad s=-1,0,1,2\qquad s+m+l=0 \mod 2
\ee
and then by hand compensate for overcounting.
We then label the fields by $\Phi^l_{m,s;\bar m,\bar s}$, with weights and charge given by 
\be\label{hlms}
h^l_{m,s}=\frac{l(l+2)-m^2}{4(k+2)}+\frac{s^2}8 \mod 1\ ,
\qquad Q^l_{m,s}=\frac{m}{k+2}-\frac s2 \mod 2\ .
\ee
Their characters are given by \cite{Gepner:1987qi}
\be
\chi_m^{l(s)}(\tau,z)= \sum_{j\mod k} c^l_{m+4j-s}(\tau) \Theta_{2m+(4j-s)(k+2),2k(k+2)}(\tau,2kz,0)\ .
\ee
Here the classical theta function associated with $SU(2)$ at level $m$ is
\be
\Theta_{n,m}(\tau,z,u)= e^{-2\pi i u} \sum_{j\in \Z+\frac n{2m}} e^{2\pi \tau m j^2+2\pi i jz}
\ee
and $c^l_{m}$ is the string function of $A^{(1)}_1$,
\be
c^l_m(\tau) = \eta(\tau)^{-3} \sum_{\substack{-|x|<y\leq|x|\\ (x,y) \textrm{ or } (1/2-x,1/2+y)\in (\frac{l+1}{2(k+2)}, \frac{m}{2k})+\Z^2}} sign(x) e^{2\pi i\tau((k+2)x^2-ky^2}\ .
\ee
These characters are indeed invariant under $s\mapsto s +4$, $m \mapsto m + 2k+4$ and under the identification $l\mapsto k-l, m\mapsto m+k+2, s\mapsto s+2$.

\subsection{The $k=2$ minimal model and correlation functions}\label{ss:k2Ising}
The correlation functions of minimal models can be obtained by using the coset construction and then relating them to the $su(2)$ WZW correlation functions originally derived in \cite{Zamolodchikov:1986bd} --- see  \cite{Mussardo:1988ck,Mussardo:1988av} for explicit expressions. We will instead use the parafermion description \cite{Fateev:1985mm,Zamolodchikov:1986gh}. We will focus on the $k=2$ minimal model and use the fact that the $\Z_2$ parafermion theory is simply the Ising model, that is the $c=1/2$ Virasoro minimal model. We can thus write the $k=2$ $N=2$ minimal model as the tensor product of the Ising model with a free boson at radius $\sqrt{2}$, and obtain all correlation functions from Ising and free boson correlation functions.

Let us give a brief reminder on the Ising model --- see \eg \cite{DiFrancesco:1997nk,Ginsparg:1988ui} for a more detailed review. The Ising model has operators $\psi,\bar \psi, \sigma, \mu$ with weights
\be
(h,\bar h)_\sigma = (h,\bar h)_\mu = (1/16,1/16)\ , \qquad (h,\bar h)_\psi =(1/2,0)\ ,
\qquad (h,\bar h)_{\bar \psi} =(0,1/2)\ .
\ee
We will use the following correlation functions for $\psi$ and $\bar \psi$   \cite{DiFrancesco:1997nk}:
\be
F_{\psi\bar\psi}(x,y)=\langle \sigma(\infty)\psi(x)\bar\psi(\bar y) \sigma(0) \rangle = \frac1{2x^{1/2}\bar y^{1/2}}
\ee
and
\be
F_{\psi\psi}(x,y)=\frac1{2(x-y)}\left(\sqrt{\frac xy}+\sqrt{\frac yx}\right) 
\ee
and
\be
F_{\psi\bar\psi\psi\bar\psi}(x,y)=\langle \sigma(\infty)\psi(x)\bar\psi(\bar x)\psi(y)\bar\psi(\bar y) \sigma(0) \rangle= \frac1{4|x-y|^2}\left(\sqrt{\frac xy}+\sqrt{\frac yx}\right)\left(\sqrt{\frac{ \bar x}{\bar y}}+\sqrt{\frac {\bar y}{\bar x}}\right)\ .
\ee
The correlation functions for $\sigma$ and $\mu$ can be obtained by using the null vectors of the Ising model \cite{Ginsparg:1988ui}. We will use
\begin{multline}
F_{\sigma\sigma\sigma\sigma}(x,\bar x) =
\langle \sigma(\infty) \sigma(1,1) \sigma(x,\bar x) \sigma(0)\rangle
=F_{\mu\mu\mu\mu}(x,\bar x)
= \langle \mu(\infty) \mu(1,1) \mu(x,\bar x) \mu(0)\rangle\\
= \frac{1}{\sqrt 2 |x|^{1/4}|1-x|^{1/4}}\sqrt{1+|x|+|1-x|} 
= \frac{1}{2 |x|^{1/4}|1-x|^{1/4}}\left(|1+\sqrt{1-x}|+|1-\sqrt{1-x}|
\right)
\end{multline}
and
\begin{multline}
F_{\sigma\mu\mu\sigma}(x,\bar x) = \langle \sigma(\infty) \mu(1,1) \mu(x,\bar x) \sigma(0)\rangle =
\\
\frac{1}{2 |x|^{1/4}|1-x|^{1/4}}\left((1-\sqrt{1-x})^{1/2}(1+\sqrt{1-\bar x})^{1/2} + (1+\sqrt{1-x})^{1/2}(1-\sqrt{1-\bar x})^{1/2}
\right)\ .
\end{multline}
In fact we will see that it is more useful to work with the correlator after the transformation $z\mapsto 1-z$, namely
\begin{multline}
F_{\sigma\sigma\mu\mu}(x,\bar x) = \langle \sigma(\infty) \sigma(1,1) \mu(x,\bar x) \mu(0)\rangle =
\\
 \frac{1}{2 |x|^{1/4}|1-x|^{1/4}}\left((1-\sqrt{x})^{1/2}(1+\sqrt{\bar x})^{1/2} + (1+\sqrt{x})^{1/2}(1-\sqrt{\bar x})^{1/2}
\right)\ .
\end{multline}
As mentioned above, these correlation functions can be obtained by computing the Virasoro conformal blocks for $c=1/2$ by solving the differential equation for the null vector, and then fixing the coefficients in their sesquilinear combination by matching monodromies and factorization into two point functions. For instance, to find $F_{\sigma\sigma\mu\mu}$ we used $F_{\sigma\sigma\mu\mu}(x,\bar x) \sim |x|^{-1/4} = \langle \sigma(\infty) \sigma(1,1)\rangle \langle\mu(x,\bar x) \mu(0)\rangle$ for $x\to 0$ and the fact that there is no monodromy as $x$ circles around 0. We do however pick up a minus sign when $x$ circles around $\infty$. To see this, it is useful to write the correlator as
\be
F_{\sigma\sigma\mu\mu}(x,\bar x)= \frac{i}{2 |x|^{1/4}|1-x|^{1/4}}\left((\sqrt{x}-1)^{1/2}(\sqrt{\bar x}+1)^{1/2} - (\sqrt{x}+1)^{1/2}(\sqrt{\bar x}-1)^{1/2}
\right)
\ee
and then consider the limit $x\to \infty$; this gives indeed the expected branch cut at infinity.

For the correlation functions of the free boson part we have the usual expression
\be\langle\prod_i :e^{i k_i \phi(z_i)}::e^{i \bar k_i \bar\phi(z_i)}: \rangle\ 
= \prod_{i<j}(z_i-z_j)^{k_ik_j}(\bar z_i-\bar z_j)^{\bar k_i\bar k_j}\ .
\ee
For future convenience we define the holomorphic part of the correlator of four free bosons at $0,1,x,\infty$ in terms of rescaled momenta
\be\label{BosCorr}
B_{l_2l_3l_4}(x) :=  x^{l_3l_4/8}(1-x)^{l_2l_3/8}\ ,
\ee
and similarly for the anti-holomorphic part.

Finally, to make the connection to the minimal model correlators explicit, let us give the dictionary between the fields of the $k=2$ minimal model and its Ising realization  \cite{EberlePhD}. The symmetry algebra is realized as\footnote{Note that our normalization of the $G$ differs from  \cite{EberlePhD} by a factor of $\sqrt 2$.}
\be
G^-(z) = \frac1{\sqrt2}\psi(z) e^{i\sqrt{2}\phi(z)}\ ,
\qquad G^+(z) = \frac1{\sqrt2}\psi(z) e^{-i\sqrt{2}\phi(z)}\ ,
\ee
and
\be
J(z) = -\frac{i}{\sqrt{2}}\partial \phi(z)\ ,
\ee
so that the field $e^{ik\phi(z)}$ has $h=k^2/2$ and $Q=-k/\sqrt2$. The anti-holomorphic symmetry algebra has similar expressions.
Moreover we need the primary fields
\be
N^1_{\pm1,\pm1} = \sigma(z,\bar z) e^{\mp \frac{i}{2\sqrt{2}}\phi(z)} e^{\mp \frac{i}{2\sqrt{2}}\bar\phi(\bar z)} \ ,
\qquad
N^1_{\pm1,\mp1} = \mu(z,\bar z) e^{\mp \frac{i}{2\sqrt{2}}\phi(z)} e^{\pm \frac{i}{2\sqrt{2}}\bar\phi(\bar z)} \ ,
\ee
and
\be
N^2_{\pm2,\pm2} = e^{\mp\frac{i}{\sqrt{2}}\phi(z)} 
e^{\mp \frac{i}{\sqrt{2}}\bar\phi(\bar z)}\ .
\ee
We also need some of their $G$ descendants, namely
\begin{multline}
G^-_{-1/2}N^2_2(z)= \oint_z dw G^-(w) N^2_2(z) = \oint_z dw \frac1{\sqrt2}\psi(w) :e^{i\sqrt{2}\phi(w)}: :e^{\frac{-i} {\sqrt{2}}\phi(z)}: \\
 = \oint_z dw \frac1{\sqrt2}\psi(w) (w-z)^{-1} :e^{\frac i {\sqrt{2}}\phi(z)}:+\ldots
 = \frac1{\sqrt2}\psi(z) :e^{\frac i {\sqrt{2}}\phi(z)}:
\end{multline}
and
\begin{multline}
G^-_{-1/2}N^1_{1,1}(z, \bar z)= 
%\oint_z dw G^-(w) N^1_1(z) = 
\oint_z dw \frac1{\sqrt2}\psi(w) :e^{i\sqrt{2}\phi(w)}: \sigma(z,\bar z):e^{\frac{- i} {2\sqrt{2}}\phi(z)}: \\
 = \oint_z dw  \frac{e^{i\pi/4}}{2}(w-z)^{-1} :e^{\frac {3i} {2\sqrt{2}}\phi(z)}: \mu(z,\bar z)e^{\frac{-i}{2\sqrt{2}}\bar\phi(\bar z)} +\ldots
 = \frac{e^{i\pi/4}}{2}  \mu(z,\bar z) e^{\frac {3i} {2\sqrt{2}}\phi(z)} e^{\frac{-i}{2\sqrt{2}}\bar\phi(\bar z)} \ .
\end{multline}
Here we used the OPE of $\psi$ with $\sigma$, see for instance \cite{DiFrancesco:1997nk}.
Similarly we have
\bea
\bar G^-_{-1/2}N^1_{1,1}(z, \bar z) &=& \frac{e^{-i\pi/4}}{{2}}  \mu(z,\bar z)  e^{ \frac{-i}{2\sqrt{2}}\phi(z)} e^{\frac {3i} {2\sqrt{2}}\bar \phi(z)}\ ,\\
G^-_{-1/2}N^1_{1,-1}(z, \bar z)&=& \frac{e^{-i\pi/4}}{{2}}  \sigma(z,\bar z) e^{\frac {3i} {2\sqrt{2}}\phi(z)} e^{\frac{i}{2\sqrt{2}}\bar\phi(\bar z)}\ ,\\
\bar G^-_{-1/2}N^1_{-1,1}(z, \bar z) &=& \frac{e^{i\pi/4}}{{2}}  \sigma(z,\bar z)  e^{ \frac{i}{2\sqrt{2}}\phi(z)} e^{\frac {3i} {2\sqrt{2}}\bar \phi(z)}\ .
\eea
The expressions for $G^+_{-1/2}$ are the same up to the change in the charge.

\section{Gepner models}
\subsection{The Quartic Gepner model: $(2)^4$}

We can now use these $\cN=2$ minimal models to construct a Gepner model. Gepner models were introduced in \cite{Gepner:1987qi,Gepner:1989gr}.  Their basic ingredient is a tensor product of $r$ minimal models such that their total central charge satisfies
\be
\sum_{i=1}^r \frac{3k_i}{k_i+2} = 3D\ .
\ee
Here $D$ is the complex dimension of the Calabi-Yau $D$-fold whose sigma model they represent, in our case  $D=2$ for K3. More concretely, we will focus on the Gepner model with $r=4, k_i=2$. This corresponds to the quartic Fermat surface.

The idea is then to construct the K3 sigma model partition function out of the characters of this tensor product. The natural starting point is simply the diagonal invariant of the tensor product of the bosonic subalgebra of the factors,
\be
Z= \sum_{(l_i,m_i,s_i)} \prod_{i=1}^4 \chi_{m_i}^{l_i(s_i)} (\chi_{m_i}^{l_i(s_i)})^*\ .
\ee
This partition function is clearly modular invariant, as it is a product of modular invariant functions. 
It does not however describe the CY sigma model that we are interested in. 
On the one hand, there is no proper notion of Ramond and NS sectors yet: in the tensor product, the individual factors are independently allowed to be in NS and Ramond sectors.
On the other hand, it does not contain the holomorphic top form $\Omega^{D/2,0}$ that acts as the (one unit) spectral flow operator. To construct the partition function of a consistent CFT, we want to perform a so-called simple current extension \cite{Schellekens:1989am,Schellekens:1990xy,Fuchs:1996dd}. The advantage of this relatively technical way of constructing Gepner models is that it not only guarantees a modular invariant partition function, but also allows to compute (consistent) correlators of the resulting theory.

As a side remark we note that strictly speaking we do not construct a Gepner extension, but rather what \cite{Fuchs:2000gv} call a Calabi-Yau extension. The resulting theory describes the CY sigma model rather than the worldsheet theory of the string compactification. The main difference is that we extend by the one-unit spectral flow operator rather than the half-unit operator. This means that the resulting theory only has NS-NS and R-R sectors and none of the NS-R sectors of the string worldsheet theory that lead to spacetime fermions and supersymmetry. The two extensions are closely related however and it is straightforward to go from the CY extension to the Gepner extension.

Simple current extensions work in the following way:
A simple current $\Jsf$ is a holomorphic field whose fusion with any primary field $\varphi$ yields just a single field $\Jsf\varphi$. Single currents form a group $\cG$. We want to extend the chiral algebra of our CFT by such a $\Jsf$. To do this, we first project out any fields $\varphi$ whose monodromy charge
\be
Q_\Jsf(\varphi) := h_{\Jsf}+h_{\varphi}- h_{\Jsf \varphi}
\ee
is not integral. This ensures that $\Jsf$ is local, that is that the OPE of $\Jsf$ with $\varphi$ does not lead to branch cuts. The primary fields $\varphi$ that survive the projection are then organized into orbits $[\varphi ]$ of $\cG$. The diagonal modular invariant of the extended theory is
\be\label{Zext}
Z_{ext} = \sum_{\substack{[\varphi] \\ Q_\Jsf(\varphi)\in\Z\ \forall \Jsf\in\cG}}
|\cS_\varphi| \left| \sum_{\Jsf\in \cG/\cS_\varphi} \chi_{\Jsf\varphi}(\tau)\right|^2\ ,
\ee
where the $\chi$ are the characters of the original theory.
Here $\cS_\varphi$ is the subgroup of $\cG$ that leaves $\varphi$ invariant under the fusion product; we will return to it momentarily. Note that (\ref{Zext}) is non-diagonal with respect to the original characters. Off-diagonal states are often called \emph{twisted} sectors of the extension, even though they are strictly speaking not the same as twisted sectors of orbifold theories.

Let us now apply this to our minimal models.
We first want to extend the model by simple currents written schematically as $G^i G^j$. Here the indices denote in which factor the current lives, such as
\be
G^1G^2=\Phi^0_{0,2;0,0}\otimes\Phi^0_{0,2;0,0}\otimes\Phi^0_{0,0;0,0}\otimes\Phi^0_{0,0;0,0}\ .
\ee
These form a group $\Z_2^3$. From (\ref{hlms}) we see that their projection implements $s_i/2+s_j/2\in\Z$, meaning that all factors are either in the NS or Ramond sector.
Including their twisted  sectors leads to a partition function that is a sum over terms
\be
\prod_{i=1}^4 \chi^{l_i(s_i)}_{m_i}(\chi^{l_i(\bar s_i)}_{m_i})^*\ ,
\ee
where the $s_i$ and $\bar s_i$ are all either odd or even (for the R-R or NS-NS sector respectively) and satisfy the GSO-type condition
\be\label{GSO}
\sum_{i=1}^4 \left(\frac{s_i}2+\frac{\bar s_i}2 \right) \in 2\Z\ .
\ee
Next we want to extend by the one-unit spectral flow operator $\Jsf$.
This operator is given by 
\be
\Jsf= \bigotimes^4 \Phi^0_{2,2;0,0} =  \prod_i e^{\frac i{\sqrt2}\phi_i(z)}\ ,
\ee
which indeed has $h=1$ and $Q=2$. It generates a group $\Z_4$. Following the notation in \cite{Recknagel:1997sb, Brunner:2006tc}, we label the twisted sectors that arise from this extension by an integer $n=0,1,2,3$. The character of the twist $n$ sector is then given by
\be\label{twistchar}
\prod_{i=1}^4 \chi^{l_i(s_i)}_{m_i+n}(\chi^{l_i(\bar s_i)}_{m_i-n})^*\ ,
\ee
where the $s_i,\bar s_i$ satisfy the same conditions as above. In addition, it is straightforward to see that the projection condition for $\Jsf$ is
\be\label{mchargeintegral}
\sum_{i=1}^4 \frac{m_i}4 \in \Z\ .
\ee
In total, the partition function of the K3 sigma model at the Gepner point $(2)^4$ is given by summing (\ref{twistchar}) over $l_i,m_i,s_i,\bar s_i,n$ subject to the conditions (\ref{labelcond}), (\ref{GSO}), (\ref{mchargeintegral})
and the identification (\ref{fieldid}).

\subsection{Expressions for the characters}\label{ss:charexp}
For concreteness, let us give the first few terms of the partition function in the NS-NS sector of our quartic Gepner models. This will in particular allow us to identify the chiral primaries of the theory and its lightest non-BPS states. 
For the $n=0$ sector we get
\begin{multline}
Z_{n=0}=1 +19 \sqrt{q} y \sqrt{\bar{q}} \bar{y}+\frac{19 \sqrt{q} \sqrt{\bar{q}}}{y \bar{y}}+q y^2 \bar{q} \bar{y}^2+\frac{q \bar{q}}{y^2 \bar{y}^2}
\\
+12 q^{1/4} \bar{q}^{1/4}+4 \bar{q}+46 \sqrt{q} \sqrt{\bar{q}}+4 q 
\\
+24 q^{3/4} y \bar{q}^{3/4} \bar{y}+\frac{12 q^{3/4} \bar{q}^{3/4} \bar{y}}{y}+\frac{12 q^{3/4} y \bar{q}^{3/4}}{\bar{y}}+\frac{24 q^{3/4} \bar{q}^{3/4}}{y \bar{y}}+48 q^{3/4} \bar{q}^{3/4}
\\
+128 q y \bar{q} \bar{y}
+\frac{88 q \bar{q} \bar{y}}{y}+\frac{88 q y \bar{q}}{\bar{y}}+\frac{128 q \bar{q}}{y \bar{y}}+372 q \bar{q}+\ldots \ .
\end{multline}
For $n=1$ we get
\begin{multline}
Z_{n=1}=\frac{\sqrt{q} y \sqrt{\bar{q}}}{\bar{y}}+\frac{19 \sqrt{q} \sqrt{\bar{q}} \bar{y}}{y} 
+\frac{q}{y^2}+\bar{q} \bar{y}^2\\
+18 \sqrt{q} \sqrt{\bar{q}}
+
12 q^{3/4} y \bar{q}^{3/4} \bar{y}+\frac{24 q^{3/4} \bar{q}^{3/4} \bar{y}}{y}+\frac{12 q^{3/4} \bar{q}^{3/4}}{y \bar{y}}+48 q^{3/4} \bar{q}^{3/4}
\\
+\frac{4 q \bar{q}}{y^2}+4 q \bar{q} \bar{y}^2+88 q y \bar{q} \bar{y}+\frac{128 q \bar{q} \bar{y}}{y}+\frac{48 q y \bar{q}}{\bar{y}}+\frac{88 q \bar{q}}{y \bar{y}}+284 q \bar{q}+\ldots \ .
\end{multline}
For $n=2$ we get
\begin{multline}
Z_{n=2}=\sqrt{q} y \sqrt{\bar{q}} \bar{y}+\frac{\sqrt{q} \sqrt{\bar{q}}}{y \bar{y}}+\frac{q \bar{q} \bar{y}^2}{y^2}+\frac{q y^2 \bar{q}}{\bar{y}^2}\\
+6 \sqrt{q} \sqrt{\bar{q}}+
\frac{12 q^{3/4} \bar{q}^{3/4} \bar{y}}{y}+\frac{12 q^{3/4} y \bar{q}^{3/4}}{\bar{y}}+48 q^{3/4} \bar{q}^{3/4}\\
+48 q y \bar{q} \bar{y}+\frac{88 q \bar{q} \bar{y}}{y}+\frac{88 q y \bar{q}}{\bar{y}}+\frac{48 q \bar{q}}{y \bar{y}}+212 q \bar{q}+\ldots\ .
\end{multline}
For $n=3$ we get
\begin{multline}
Z_{n=3}=\frac{\sqrt{q} \sqrt{\bar{q}} \bar{y}}{y}+\frac{19 \sqrt{q} y \sqrt{\bar{q}}}{\bar{y}}+ \frac{\bar{q}}{\bar{y}^2}+q y^2
\\
+18 \sqrt{q} \sqrt{\bar{q}}+12 q^{3/4} y \bar{q}^{3/4} \bar{y}+\frac{24 q^{3/4} y \bar{q}^{3/4}}{\bar{y}}+\frac{12 q^{3/4} \bar{q}^{3/4}}{y \bar{y}}+48 q^{3/4} \bar{q}^{3/4}
\\
+4 q y^2 \bar{q}+88 q y \bar{q} \bar{y}+\frac{48 q \bar{q} \bar{y}}{y}+\frac{128 q y \bar{q}}{\bar{y}}+\frac{88 q \bar{q}}{y \bar{y}}+\frac{4 q \bar{q}}{\bar{y}^2}+284 q \bar{q}+\ldots \ .
\end{multline}
Collecting the four sectors, we see that in terms of BPS states, we indeed get 80 moduli, the vacuum, 4 $(\pm2,\pm2)$ chiral primaries and 4 $(0,\pm2)$ and $(\pm2,0)$ chiral primaries. As for the lightest non-BPS states, we find 12 uncharged $(1/4,1/4)$ states, all of them in the $n=0$ sector. These are the states whose lifting we will analyze.

\subsection{Chiral primaries}

Let us describe the moduli that we found in the character in more detail.
The $k=2$ minimal model has chiral fields $N^l_l$ with $h=l/8$ and $q=l/4$, and anti-chiral fields $N^l_{-l}$ with $h=l/8$ and $q=-l/4$, with $l=0,1,2$.

For the untwisted $n=0$ sector, the main constraint that determines the number of chiral fields comes from (\ref{mchargeintegral}). The sum is simply the total $U(1)$ charge of the state. The only cc field that gives $Q=0$ is of course the vacuum $(N^0_{0;0})^{\otimes4}$. There is also one cc field with $Q=2$, namely $(N^2_{2;2})^{\otimes 4}$. For $Q=1$, the chiral primaries are products of factors of $N^l_{l;l}$ such that the total $U(1)$ charge is $Q=1$. 
Up to permutations, there are three possible ways to construct such primaries: four factors of $N^1_{1;1}$ (which we will call type A), two factors $N^2_{2;2}$ (type B), and one factor $N^2_{2;2}$ and two factors $N^1_{1;1}$ (type C). A quick combinatorial argument shows that there are 1, 6 and 12 fields of the respective types.
In summary, we have the following moduli:
\bea1 \textrm{ type } A &:& N^1_{1;1}\otimes N^1_{1;1}\otimes N^1_{1;1}\otimes N^1_{1;1}\\
6 \textrm{ type } B &:& N^2_{2;2}\otimes N^2_{2;2}\otimes 1\otimes 1  \qquad \textrm{(and permutations)}\\
12 \textrm{ type } C &:& N^2_{2;2}\otimes N^1_{1;1}\otimes N^1_{1;1}\otimes 1  \qquad \textrm{(and permutations)}
\qquad
\eea
Together with the $Q=1$ cc field from the $n=2$ sector found below this gives the 20 $\cN=4$ moduli of K3. The aa, ac, ca components of these moduli can be constructed in a similar way, giving the total 80 $\cN=2$ moduli of K3.

In the twisted sectors we have cc primaries in the characters
\be
(\chi^{n-1(0)}_{n-1} \chi^{n-1(-2)*}_{-n-1})^{4}\ .
\ee
For the $n=2$ twisted sector this means we have another type A modulus
\be
N^1_{1;1}\otimes N^1_{1;1}\otimes N^1_{1;1}\otimes N^1_{1;1}
\ee
with charge $(1,1)$. 
For $n=1$ we have
\be
(N^0_0)^{\otimes 4}\otimes (\bar N^2_2)^{\otimes 4}
\ee
with charge $(0,2)$ and for $n=3$
\be
(N^2_2)^{\otimes 4}\otimes (\bar N^0_0)^{\otimes 4}
\ee
with charge $(2,0)$, corresponding to the corners of the Hodge diamond.

In listing all moduli, we encounter an apparent puzzle: it seems that the type A modulus appears twice, once in the $n=0$ sector and once in the $n=2$ sector. Even though this may not be a problem on the level of the partition function, clearly those two primaries must be distinguished by their fusion rules and correlation functions. In fact, they are distinguished by an additional quantum number. This quantum number arises from the fact that the type A modulus is invariant under $\Jsf^2G^1G^2G^3G^4$ because of the field identification (\ref{fieldid}). Its stabilizer group $\cS_\varphi$ is $\Z_2$, which leads to a two-fold degeneracy that is lifted by introducing a $\Z_2$ quantum number corresponding to the two irreducible characters of $\Z_2$. In principle this affects their fusion rules --- see \eg equation (2.10) in \cite{Fuchs:2000gv}. For our purposes however this will not matter, as the correlators we are interested in vanish in either case.

\subsection{CY extension and correlation functions}
Let us briefly discuss how to compute correlation function of the extended theory. 
The correlation functions of the underlying tensor theory follow of course directly from the minimal model theory. However, for the extended theory in principle we need to employ the Verlinde formula to obtain the new fusion rules, and then compute the correlators from the conformal blocks of the bosonic subalgebra.
For the extension by $G_i G_k$, we will circumvent this procedure by working with the full $N=2$ correlation functions in the NS sector. All $G_iG_k$ twisted states are then already included. The only issue is to deal with the GSO projection when computing correlation functions. In general what we need to do is to impose the GSO projection on the internal fields. The reason for this is that the odd fusion rules can change the total fermion parity. More concretely, even fusion channels leave the fermion parity invariant. So do odd channels as long as both left- and right-movers are in an odd channel. Thus in our setup the only case that is affected is the fusion of something like $G^-_{-1/2}N^1_{1,1}(x) N^1_{1,1}(0)$, in which case the left movers are in the odd channel and the right movers are in the even channel, which will be projected out. For our calculations it turns out that this never becomes necessary. One quick way of checking this is to observe that no branch cuts appear in our correlators.

For the extension by $\Jsf$, we first note that the integral charge condition is preserved under fusion, meaning that as long as the external fields satisfy it, the internal fields do so too, so that no additional projection is needed. Next, in principle we need to include $\Jsf$ twisted states. First note that these only involve the bosonic part of the correlator. The bosonic correlation functions however are already constrained by bosonic charge conservation. For fixed external fields, charge conservation uniquely fixes the internal channel, so that no additional twisted states are allowed to appear.

\section{Lifting the $h=1/4, \bar h =1/4$ fields under type B moduli}

\subsection{First order lifting}
From the expressions in section~\ref{ss:charexp},
we see that the lightest non-BPS states have $h=1/4, \bar h=1/4$. There are 12 of them, and they all come from the $n=0$ sector. They are of the form
\be
\varphi_{12}=N^1_{1;1} \otimes N^1_{-1;-1}\otimes N^0_{0;0} \otimes N^0_{0;0}
\ee
and permutations thereof. We will denote them by $\varphi_{ij}$, where $i$ labels the factor with $N^1_{1;1}$ and $j$ the factor with $N^1_{-1;-1}$.

The types of chiral primaries of appropriate charge and weight were described above. The actual moduli are given by taking their $G$ descendants such as 
\be
O = G_{-1/2}\bar G_{-1/2}\left( N^2_{2;2}\otimes N^2_{2;2}\otimes 1\otimes 1\right)\ .
\ee
Here $G$ denotes the diagonal supercharge of all the tensor factor supercharges, that is
\be\label{Gdiag}
G\otimes 1 \otimes 1 \otimes 1+1\otimes G \otimes 1 \otimes 1 + 1\otimes 1 \otimes G \otimes 1+1\otimes 1 \otimes 1 \otimes G \ .
\ee
In principle $G=G^++G^-$ is the $N=1$ supercharge, but when acting on (anti-)chiral primaries only one of the $N=2$ supercharges survives. Note that even though $G$ is not in the theory due to the GSO projection, $G\bar G$ is. Finally we need to define a hermitian modulus, which we do as
\be\label{Phihermitian}
\Phi = \frac{O+O^\dagger}{\sqrt 2}\ .
\ee
The alternative hermitian linear combination  $i(O-O^\dagger)/\sqrt2$ leads to a vanishing first order contribution, as can be seen from the expressions below.\footnote{We thank an anonymous referee for pointing this out.} Since we are looking for non-vanishing contributions, we will focus on (\ref{Phihermitian}).

Let us first discuss the first order lifting for the type A modulus. Since there are 12 fields $\varphi_{ij}$ of the same weight, we need to use degenerate perturbation theory. To obtain the entries of the lifting matrix the fusion rules (\ref{fusion1})--(\ref{fusionlast}) are useful. We immediately see that all diagonal entries vanish, since two factors will be one-point functions. A slightly more careful analysis shows that in fact all entries vanish: the only way to obtain the vacuum is the fusion $N^1_{1;1} \times N^1_{-1,-1}$, but there are only two fields $N^1_{-1;-1}$ available. The type A modulus thus does not lift the $\varphi_{ij}$ at first order. We will therefore not try to analyze its second order effects.
The same argument shows that the $n=2$ twisted modulus also gives no first order contribution.

Let us therefore turn to type B moduli. For concreteness, we will pick
\be\label{Odef}
O = G_{-1/2}\bar G_{-1/2}\left( N^2_{2;2}\otimes N^2_{2;2}\otimes 1\otimes 1\right)\ .
\ee
Any other moduli of type B will lead to the same result after permuting factors.

Let us first discuss the form of the first order lifting matrix $\langle\varphi_i^\dagger \Phi \varphi_j\rangle$. Note that even though $O$ has overall charge zero, the first two factors have charge $\pm \frac12$. This means that the only non-vanishing entries occur between fields $\varphi_i$ and $\varphi_j$ whose charges in the first and sector factor differ by exactly one. This implies in particular that the diagonal elements of the matrix vanish. In fact, it follows that the only non-vanishing entries occur between the two fields
\be
\varphi_{12}:=N^1_{1;1} \otimes N^1_{-1;-1}\otimes 1 \otimes 1 \ ,
\qquad \varphi_{21}:=N^1_{-1;-1} \otimes N^1_{1;1}\otimes 1 \otimes 1 \ ,
\ee
which satisfy $\varphi_{12}^\dagger =\varphi_{21}$.

Now we can compute the 3-pt function $\langle \varphi_{12} O \varphi_{12}\rangle$ using the results in section~\ref{ss:k2Ising}. We have two non-trivial factors, each of which we consists of an Ising part and a free boson part:\be
\langle N^1_{-1;-1} N^1_{-1;-1} N^2_{2;2}\rangle = \langle \sigma \sigma \rangle  =1
\ee
and
\be
\langle N^1_{1;1} N^1_{1;1} G^-_{-1/2} \bar G^-_{-1/2} N^2_{2;2}\rangle
=\frac12\langle \sigma\sigma \psi\bar\psi\rangle = \frac12F_{\psi\bar\psi}(1,1)=\frac14\ .
\ee
Similarly, $\langle \varphi_{12} O^\dagger \varphi_{12}\rangle=\frac14$. In total we thus have
\be\label{B3pt}
C_{\varphi_{21}^\dagger \Phi \varphi_{12}}= \frac1{2\sqrt 2}
\ee
and similarly for $C_{\varphi_{12}^\dagger \Phi \varphi_{21}}$.
Diagonalizing the matrix gives two eigenstates 
\be
\varphi_\pm=\frac{\varphi_{12}\pm \varphi_{21}}{\sqrt 2}
\ee
that get lifted by 
\be
h^{(1)} = \mp  \frac \pi { 2\sqrt 2}\ .
\ee

\subsection{Second order lifting}
Let us now compute the lifting of the states $\varphi_\pm$ at second order. Since those two states are non-degenerate after perturbation at first order, at second order we can use non-degenerate perturbation theory using the correlator 
\be
\langle \varphi_\pm^\dagger(\infty) \Phi(1)\Phi(x) \varphi_\pm(0) \rangle\ .
\ee
When expressed in terms of $\varphi_{12}$ and $\varphi_{21}$, the cross terms vanish, and only diagonal entries are non-vanishing. This follows from the fusion rules (\ref{f22}) and (\ref{f2m2}). Moreover, by hermitian conjugation it is clear that $\varphi_{12}$ and $\varphi_{21}$ give the same contribution. For concreteness we will thus compute $\langle \varphi_{12}(\infty) \Phi(1)\Phi(x) \varphi_{21}(0) \rangle$.

To do this, we expand out the actual hermitian modulus $\Phi=\frac1{\sqrt{2}}(O+O^\dagger)$. This leads to four correlators with $O(1)O(x), O^\dagger(1)O^\dagger(x), O^\dagger(1)O(x)$ and $O(1)O^\dagger(x)$ inserted that need to be evaluated. Let us discuss them one by one.

\subsubsection{ $O(1)O(x)$}
When working out the diagonal $G_{-1/2}$ operator (\ref{Gdiag}), the $G_{-1/2}$ annihilate the vacua in the last two factors, so that each $G$ and $\bar G$ only leads to 2 terms. The two moduli thus lead to $2^4=16$ terms in total. However, factorwise charge conservation makes most of these terms vanish: each factor has to contain exactly one $G^-$ and one $\bar G^-$, meaning that once we have distributed the $G$ of $O(1)$ to the two factors, for which there are 4 possibilities, the distribution of the $G$ of $O(x)$ is completely fixed. There are thus four configurations, which we evaluate in turn.

\begin{enumerate}
\item
Take the first factor to be of the form
\be
\langle N^1_{1;1}(\infty) G^-_{-1/2} \bar G^-_{-1/2} N^2_{2;2}(1) N^2_{2;2}(x)N^1_{-1;-1}(0)\rangle\ .
\ee
This correlation function can be written as the product of bosonic part and an Ising part. 
In the notation of (\ref{BosCorr}), the bosonic part is
\be
B_{2-21}(x) B_{2-21}(\bar x) = x^{-1/4}(1-x)^{-1/2}\bar x^{-1/4}(1-\bar x)^{-1/2}\ .
\ee
The Ising part is given by
\be
\frac12 F_{\psi\bar \psi}(1,1)=\frac14\ ,
\ee
so that in total we get
\be
\frac14 |x|^{-1/2}|1-x|^{-1}
\ee
As a sanity check, we note that the pole $x^{-1/4}$ comes from the fusion $N^2_2(x)N^1_{-1}(0) \sim x^{-1/4}N^1_1(0)$, and the pole $(1-x)^{-1/2}$ comes from $(G^-_{-1/2}N^2_2)(1)N^2_2(x) \sim (1-x)^{-1/2}N^2_0(x)$

The second factor necessarily has to have both $G$ from $O(x)$. It is therefore of the form
\be  
\langle N^1_{-1;-1}(\infty)N^2_{2;2}(1)  G^-_{-1/2} \bar G^-_{-1/2} N^2_{2;2}(x)N^1_{1;1}(0)\rangle\ .
\ee
The bosonic part is now
\be
B_{-22-1}(x) B_{-22-1}(\bar x) = x^{-1/4}(1-x)^{-1/2}\bar x^{-1/4}(1-\bar x)^{-1/2}\ ,
\ee
and the Ising part is
\be
\frac12 F_{\psi\bar \psi}(x,\bar x)=\frac14 x^{-1/2}\bar x^{-1/2}\ ,
\ee
giving a total of
\be
\frac14 |x|^{-3/2}|1-x|^{-1}\ .
\ee
Now the pole $x^{-3/4}$ comes from the fusion $(G^-_{-1/2}N^2_2)(x)N^1_{1}(0) \sim x^{-3/4}N^1_{-1}(0)$, and the pole $(1-x)^{-1/2}$ is the same as above.

The third and fourth factors are of course trivial. Altogether the contribution from this configuration is thus
\be
\frac1{16} |x|^{-2}|1-x|^{-2}\ .
\ee

\item
The next configuration has a first factor of
\be  
\langle N^1_{1;1}(\infty)N^2_{2;2}(1)  G^-_{-1/2} \bar G^-_{-1/2} N^2_{2;2}(x)N^1_{-1;-1}(0)\rangle\ .
\ee
The bosonic part is now
\be
B_{-221}(x) B_{-221}(\bar x) = x^{1/4}(1-x)^{-1/2}\bar x^{1/4}(1-\bar x)^{-1/2}\ .
\ee
and the Ising part is
\be
\frac12F_{\psi\bar \psi}(x,\bar x)=\frac14 x^{-1/2}\bar x^{-1/2}\ ,
\ee
giving
\be
\frac14 |x|^{-1/2}|1-x|^{-1}\ .
\ee
The second factor is
\be
\langle N^1_{-1;-1}(\infty) G^-_{-1/2} \bar G^-_{-1/2} N^2_{2;2}(1) N^2_{2;2}(x)N^1_{1;1}(0)\rangle
\ee
giving
\be
B_{2-2-1}(x) B_{2-2-1}(\bar x) = x^{1/4}(1-x)^{-1/2}\bar x^{1/4}(1-\bar x)^{-1/2}\ .
\ee
and
\be
\frac12 F_{\psi\bar \psi}(1,1)=\frac14\ ,
\ee
giving a total of
\be
\frac14 |x|^{1/2}|1-x|^{-1}
\ee
The total contribution is thus
\be
\frac1{16} |1-x|^{-2}\ .
\ee

\item
The first factor is
\be  
\langle N^1_{1;1}(\infty)G^-_{-1/2}N^2_{2;2}(1)   \bar G^-_{-1/2} N^2_{2;2}(x)N^1_{-1;-1}(0)\rangle
\ee
giving
\be
\frac14 \bar x^{-1/2} x^{-1/4}\bar x^{1/4}(1-x)^{-1/2} (1-\bar x)^{-1/2} = \frac14 |x|^{-1/2}|1-x|^{-1}\ .
\ee
The second factor is
\be  
\langle N^1_{-1;-1}(\infty)\bar G^-_{-1/2}N^2_{2;2}(1)   G^-_{-1/2} N^2_{2;2}(x)N^1_{1;1}(0)\rangle
\ee
giving
\be
\frac14  x^{-1/2} x^{-1/4}\bar x^{1/4}(1-x)^{-1/2} (1-\bar x)^{-1/2} = \frac14 |x|^{-3/2}\bar x |1-x|^{-1}\ .
\ee
In total we get
\be
\frac1{16} \bar x |x|^{-2}|1-x|^{-2}\ .
\ee

\item
The first factor is
\be  
\langle N^1_{1;1}(\infty)\bar G^-_{-1/2}N^2_{2;2}(1)   G^-_{-1/2} N^2_{2;2}(x)N^1_{-1;-1}(0)\rangle\ ,
\ee
giving
\be
\frac14 x^{-1/2} x^{1/4}\bar x^{-1/4}(1-x)^{-1/2} (1-\bar x)^{-1/2} = \frac14 |x|^{-1/2}|1-x|^{-1}\ .
\ee
The second factor is
\be  
\langle N^1_{-1;-1}(\infty)G^-_{-1/2}N^2_{2;2}(1)   \bar G^-_{-1/2} N^2_{2;2}(x)N^1_{1;1}(0)\rangle\ ,
\ee
giving
\be
\frac14 \bar x^{-1/2} x^{1/4}\bar x^{-1/4}(1-x)^{-1/2} (1-\bar x)^{-1/2} = \frac14 |x|^{-3/2}x |1-x|^{-1}\ .
\ee
In total we get
\be
\frac1{16} x |x|^{-2}|1-x|^{-2}\ .
\ee

\end{enumerate}

Collecting the contributions from all configurations we get
\be
\frac1{16} \frac{|1+x|^2}{|x|^2|1-x|^2}
\ee

\subsubsection{$O^\dagger(1)O^\dagger(x)$}
To compute this contribution, we can simply take the hermitian conjugate of the previous contribution. $\varphi_{21}$ and $\varphi_{21}^\dagger$ only differ by switching the first two factors. The result is thus the same as above,
\be
\frac1{16} \frac{|1+x|^2}{|x|^2|1-x|^2}\ .
\ee

\subsubsection{$O^\dagger(1)O(x)$}
We have again four possible configurations:
\begin{enumerate}
\item
The first factor is
\be
\langle N^1_{1;1}(\infty) N^2_{-2;-2}(1) N^2_{2;2}(x)N^1_{-1;-1}(0)\rangle\ .
\ee
We have 
\be
B_{2-21}(x) B_{2-21}(\bar x)
=x^{-1/4}\bar x^{-1/4}(1-x)^{-1/2}(1-\bar x)^{-1/2}
= |x|^{-1/2}|1-x|^{-1}\ .
\ee
The second factor is
\be
\langle N^1_{-1;-1}(\infty) G^+_{-1/2} \bar G^+_{-1/2} N^2_{-2;-2}(1) G^-_{-1/2} \bar G^-_{-1/2} N^2_{2;2}(x)N^1_{1;1}(0)\rangle\ ,
\ee
giving
\be
B_{-22-1}(x) B_{-22-1}(\bar x)=x^{-1/4}\bar x^{-1/4}(1-x)^{-1/2} (1-\bar x)^{-1/2} = |x|^{-1/2}|1-x|^{-1}
\ee
for the bosonic part, and
\be
\frac14F_{\psi\bar\psi\psi\bar\psi}(1,x)=\frac1{16|1-x|^2}\left(x^{1/2}+x^{-1/2}\right)\left(\bar x^{1/2}+\bar x^{-1/2}\right)
\ee
for the Ising part. 
Putting everything together, we get
\be
\frac1{16} \frac{|1+x|^2}{|x|^2|1-x|^4}\ .
\ee

\item
Next, the first factor is
\be
\langle N^1_{1;1}(\infty) G^+_{-1/2} \bar G^+_{-1/2} N^2_{-2;-2}(1) G^-_{-1/2} \bar G^-_{-1/2} N^2_{2;2}(x)N^1_{-1;-1}(0)\rangle\ .
\ee
We have 
\be
B_{-221}(x) B_{-221}(\bar x)=x^{1/4}\bar x^{1/4}(1-x)^{-1/2} (1-\bar x)^{-1/2} = |x|^{1/2}|1-x|^{-1}
\ee
for the bosonic part, and
\be
\frac14F_{\psi\bar\psi\psi\bar\psi}(1,x)=\frac1{16|1-x|^2}\left(x^{1/2}+x^{-1/2}\right)\left(\bar x^{1/2}+\bar x^{-1/2}\right)
\ee
for the Ising part. 
The second factor is 
\be
\langle N^1_{-1;-1}(\infty) N^2_{-2;-2}(1) N^2_{2;2}(x)N^1_{1;1}(0)\rangle\ .
\ee
We have 
\be
B_{2-2-1}(x) B_{2-2-1}(\bar x)=x^{1/4}\bar x^{1/4}(1-x)^{-1/2} (1-\bar x)^{-1/2} = |x|^{1/2}|1-x|^{-1}\ .
\ee
Putting everything together, we get
\be
\frac1{16} \frac{|1+x|^2}{|1-x|^4}\ .
\ee

\item
Next, the first factor is
\be
\langle N^1_{1;1}(\infty) G^+_{-1/2} N^2_{-2;-2}(1) G^-_{-1/2} N^2_{2;2}(x)N^1_{-1;-1}(0)\rangle\ .
\ee
We have 
\be
B_{-221}(x) B_{2-21}(\bar x)=x^{1/4}\bar x^{-1/4}(1-x)^{-1/2} (1-\bar x)^{-1/2}
\ee
for the bosonic part and
\be
\frac14\frac1{1-x}(x^{1/2}+x^{-1/2})
\ee
for the Ising part.
The second factor is
\be
\langle N^1_{-1;-1}(\infty) \bar G^+_{-1/2} N^2_{-2;-2}(1) \bar G^-_{-1/2} N^2_{2;2}(x)N^1_{1;1}(0)\rangle\ .
\ee
We have 
\be
B_{2-2-1}(x) B_{-22-1}(\bar x)=x^{1/4}\bar x^{-1/4}(1-x)^{-1/2} (1-\bar x)^{-1/2}
\ee
for the bosonic part and
\be
\frac14\frac1{1-\bar x}(\bar x^{1/2}+\bar x^{-1/2})
\ee
for the Ising part.
In total we thus get
\be
\frac1{16} \frac{x|1+x|^2}{|x|^2|1-x|^4}\ .
\ee

\item
Next, the first factor
\be
\langle N^1_{1;1}(\infty) \bar G^+_{-1/2} N^2_{-2;-2}(1) \bar G^-_{-1/2} N^2_{2;2}(x)N^1_{-1;-1}(0)\rangle
\ee
gives
\be
B_{2-21}(x) B_{-221}(\bar x)=
x^{-1/4}\bar x^{1/4}(1-x)^{-1/2} (1-\bar x)^{-1/2} = x^{-1/4}\bar x^{1/4}|1-x|^{-1}
\ee
for the bosonic part and
\be
\frac14\frac1{1-\bar x}(\bar x^{1/2}+\bar x^{-1/2})
\ee
for the Ising part.
The second factor is
\be
\langle N^1_{-1;-1}(\infty) G^+_{-1/2} N^2_{-2;-2}(1) G^-_{-1/2} N^2_{2;2}(x)N^1_{1;1}(0)\rangle\ .
\ee
We get
\be
B_{-22-1}(x) B_{2-2-1}(\bar x)=x^{-1/4}\bar x^{1/4}(1-x)^{-1/2} (1-\bar x)^{-1/2} = x^{-1/4}\bar x^{1/4}|1-x|^{-1}
\ee
for the bosonic part and
\be
\frac14\frac1{1-x}(x^{1/2}+x^{-1/2})
\ee
for the Ising part.
In total we thus get
\be
\frac1{16}\frac{\bar x|1+x|^2}{|x|^2|1-x|^4}\ .
\ee

\end{enumerate}
Summing this up, we get a total contribution of
\be
\frac1{16} \frac{|1+x|^4}{|x|^2|1-x|^4}\ .
\ee

\subsubsection{$O(1)O^\dagger(x)$}
By hermitian conjugation and permuting factors, this gives the same contribution as above, namely
\be
\frac1{16} \frac{|1+x|^4}{|x|^2|1-x|^4}\ .
\ee

\subsection{Integrating the overall contribution}
Collecting all contributions and accounting for the factor of $1/\sqrt{2}$ in $\Phi$, the 4-pt function is
\be
G(x)=\frac1{16} \frac{|1+x|^2}{|x|^2|1-x|^2}
+ \frac1{16} \frac{|1+x|^4}{|x|^2|1-x|^4}
= \frac{|1+x|^2(1+|x|^2)}{8|x|^2|1-x|^4}\ .
\ee
This correlation function diverges as $G(x)\sim \frac1{8|x|^2}$ at 0 and $\infty$. This is exactly as expected from (\ref{B3pt}), and will give the logarithmic divergences discussed at the end of section~\ref{ss:2ndpert}. Moreover, at $x=1$ we have $G(x) \sim |1-x|^{-4}$, which confirms that the overall normalization of the correlator is correct.

To evaluate the integral of $G(x)$, let us define the integrals
\be
I(a,b):= \int d^2 x |x|^{2a}|1-x|^{2b}\ , \label{Iint}
\ee
and
\be
J(a,b):= \int d^2 x |x|^{2a}|1-x|^{2b}(x+\bar x)\ . \label{Jint}
\ee
We thus have
\be
\int d^2 x G(x) = \frac18 (I(-1,-2)+2I(0,-2)+ I(1,-2) + J(-1,-2)+J(0,-2))\ .
\ee
To evaluate these integrals, it is useful to transform coordinates so that the integrand becomes integrable at $x=1$, meaning  $b>-1$. In particular we can use the transformation $x\mapsto 1-x$ to get the identities
\be
I(a,b)= I(b,a) \qquad J(a,b) = 2I(b,a) -J(b,a)\ 
\ee
and $x\mapsto 1/x$ to get
\be
 I(a,b)= I(-2-a-b,b) \qquad J(a,b) = J(-3-a-b,b)\ .
\ee
The total integral thus becomes
\be
\frac14(I(-2,1) + 3I(-2,0)- J(-2,0))\ .
\ee
We discuss how to evaluate integrals of the form (\ref{Iint}) and (\ref{Jint}) in appendix~\ref{app:int}. For general $a$ and $b$ they lead to hypergeometric functions. The specific values of $a$ and $b$ used here however
allow us to evaluate them in elementary form. We find
\bea
I(-2,0) &=& \frac{\pi}{\epsilon^2} + O(\epsilon^2)\ ,\\
I(-2,1) &=& \frac\pi{\epsilon^2}- 4\pi \log \epsilon + O(\epsilon^2)\ ,\\ 
J(-2,0) &=& 0\ .
\eea
This gives
\be
\frac{\pi}{\epsilon^2} - \pi \log \epsilon + O(\epsilon^2)\ .
\ee
The logarithmic divergences arise either from the square of the first order term as discussed in~\ref{ss:2ndpert}, or they arise from the singularity at $x\to 1$ and  are therefore cancelled by the counterterms we introduced in our minimial subtraction scheme.

Surprisingly, we find that in our regularization scheme the constant term vanishes, meaning that there is no second order correction to the conformal weight. Since the Gepner model has so many symmetries, this may not be altogether surprising, but it is not clear to us why exactly this happens. In any case, as we want to demonstrate the interplay between first and second order terms, we have to turn to moduli of type C.

\section{Type C moduli}
\subsection{First order lifting}
Consider now the type C modulus
\be
O = G_{-1/2}\bar G_{-1/2}\left( N^2_{2;2}\otimes N^1_{1;1}\otimes N^1_{1;1}\otimes 1\right)\ .
\ee
At first order, there is now lifting between the states
\be
\varphi_{12}:=N^1_{1;1} \otimes N^1_{-1;-1}\otimes 1 \otimes 1 
\qquad \varphi_{31}:=N^1_{-1;-1}\otimes 1 \otimes N^1_{1;1}\otimes 1 \ .
\ee
It is straightforward to see that all other matrix elements vanish.
The 3-pt function $\langle \varphi_{31}^\dagger O \varphi_{12}\rangle$ only has the non-trivial factor 
\be
\langle N^1_{1;1} N^1_{1;1} G^-_{-1/2} \bar G^-_{-1/2} N^2_{2;2}\rangle
=\frac12\langle \sigma\sigma \psi\bar\psi\rangle = \frac12 F_{\psi\bar\psi}(1,1)=\frac14\ ,
\ee
as all other factors are simply two-point functions.
On the other hand, $\langle \varphi_{31}^\dagger O^\dagger \varphi_{12}\rangle=0$ by charge conservation in the individual factors. For the modulus (\ref{Phihermitian}) we thus have in total
\be
C_{\varphi_{31}^\dagger \Phi \varphi_{12}}= \frac1{4\sqrt{2}}
\ee
and similarly for $C_{\varphi_{12}^\dagger \Phi \varphi_{13}}$.
Diagonalizing the matrix gives two eigenstates 
\be
\varphi_\pm=\frac{\varphi_{12}\pm \varphi_{31}}{\sqrt 2}
\ee
that get lifted by 
\be
h^{(1)} = \mp \pi \frac1{4\sqrt 2}\ .
\ee
We note in passing that for the alternative linear combination of $O$ and $O^\dagger$, the first and second order contributions are identical, up to a change of basis of the fields that get lifted.

\subsection{Second order lifting}
Let us now turn to second order perturbation theory and evaluate the correlator $\langle \varphi_{\pm}^\dagger \Phi\Phi \varphi_{\pm}\rangle$.
For off-diagonal entries $\langle \varphi_{12} \Phi\Phi \varphi_{31}^\dagger\rangle$ we see that the third factor has an odd number of $N^1$, so that by the fusion rules the correlator will vanish.
We can thus focus on the diagonal entries. Next we note that $\varphi_{31}^\dagger$ is actually equal to $\varphi_{12}$ up to permutation of the second and third factor. Since $\Phi^\dagger=\Phi$ and $\Phi$ is invariant under this permutation, it follows that $\langle \varphi_{12} \Phi\Phi \varphi_{12}^\dagger\rangle=\langle \varphi_{31} \Phi\Phi \varphi_{31}^\dagger\rangle$. We therefore only need to compute the matrix element $\langle \varphi_{12} \Phi\Phi \varphi_{12}^\dagger\rangle$. Just as before, the hermitian linear combination $\Phi$ leads to four different combinations of $O$ and $O^\dagger$.

\subsubsection{ $O(1)O(x)$}
From the fusion rules it follows immediately that the third factor vanishes, so the contribution is 0.

\subsubsection{ $O^\dagger(1)O^\dagger(x)$}
From the fusion rules it again follows immediately that the third factor vanishes, so the contribution is also 0.

\subsubsection{ $O(1)^\dagger O(x)$}\label{sss:OdO}

Charge conservation forces $G^-$ and $ G^+$ to be in the same factor, and same for the right-movers. In total we thus have nine configurations.

\begin{enumerate}
\item $G^-G^+$ and $\bar G^- \bar G^+$ in the first factor:

The first factor is
\be
\langle N^1_{1;1}(\infty) G^+_{-1/2} \bar G^+_{-1/2} N^2_{-2;-2}(1) G^-_{-1/2} \bar G^-_{-1/2} N^2_{2;2}(x)N^1_{-1;-1}(0)\rangle\ .
\ee
We get
\be B_{-221}(x)B_{-221}(\bar x)= |x|^{1/2}|1-x|^{-1}
\ee
for the bosonic part, and
\be
\frac14 F_{\psi\bar\psi\psi\bar\psi}(1,x)=
\frac1{16|1-x|^2}\left(x^{1/2}+x^{-1/2}\right)\left(\bar x^{1/2}+\bar x^{-1/2}\right)
= \frac{|1+x|^2}{16|1-x|^2|x|}
\ee
for the Ising part. 
The second factor is 
\be
\langle N^1_{-1;-1}(\infty) N^1_{-1;-1}(1) N^1_{1;1}(x)N^1_{1;1}(0)\rangle\ .
\ee
We have 
\be 
B_{1-1-1}(x)B_{1-1-1}(\bar x)= |x|^{1/4}|1-x|^{-1/4}
\ee
from the bosonic part and for the Ising part
\be
F_{\sigma\sigma\sigma\sigma}(x)=\frac{1}{\sqrt 2 |x|^{1/4}|1-x|^{1/4}}\sqrt{1+|x|+|1-x|} \ .
\ee
The third factor is simply
\be
|1-x|^{-1/2}\ .
\ee

Putting everything together, we get
\be
\frac1{16\sqrt2} \frac{|1+x|^2}{|x|^{1/2}|1-x|^4}\sqrt{1+|x|+|1-x|}\ .
\ee

\item $G^-G^+$ and $\bar G^- \bar G^+$ in the second factor:
The first factor is
\be
\langle N^1_{1;1}(\infty) N^2_{-2;-2}(1) N^2_{2;2}(x)N^1_{-1;-1}(0)\rangle\ ,
\ee
giving only a contribution from bosons, 
\be 
B_{2-21}(x)B_{2-21}(\bar x)
= |x|^{-1/2}|1-x|^{-1}\ .
\ee
The second factor is
\be
\langle N^1_{-1;-1}(\infty) G^+_{-1/2} \bar G^+_{-1/2} N^1_{-1;-1}(1) G^-_{-1/2} \bar G^-_{-1/2} N^1_{1;1}(x)N^1_{1;1}(0)\rangle\ .
\ee
The bosonic part is
\be
B_{-33-1}(x) B_{-33-1}(\bar x) 
\ee
and the Ising part is
\be
\frac1{16} F_{\sigma\sigma\sigma\sigma}(x,\bar x)\ ,
\ee
giving together
\be
\frac{1}{16\sqrt 2 |x||1-x|^{5/2}} \sqrt{1+|x|+|1-x|} \ .
\ee
The third factor is simply
\be
|1-x|^{-1/2}\ .
\ee
Putting everything together, we get
\be
\frac{1}{16\sqrt 2 |x|^{3/2}|1-x|^{4}} \sqrt{1+|x|+|1-x|} \ .
\ee

\item $G^-G^+$ and $\bar G^- \bar G^+$ in the third factor:
The first factor is
\be
\langle N^1_{1;1}(\infty) N^2_{-2;-2}(1) N^2_{2;2}(x)N^1_{-1;-1}(0)\rangle\ ,
\ee
giving
\be
B_{2-21}(x)B_{2-21}(\bar x)
= |x|^{-1/2}|1-x|^{-1}\ .
\ee
The second factor is 
\be
\langle N^1_{-1;-1}(\infty) N^1_{-1;-1}(1) N^1_{1;1}(x)N^1_{1;1}(0)\rangle\ ,
\ee
giving a bosonic part
\be B_{1-1-1}(x)B_{1-1-1}(\bar x)
\ee
and from the spin fields
\be
F_{\sigma\sigma\sigma\sigma}(x,\bar x)\ .
\ee
The third factor is simply
\be
\frac1{16}|1-x|^{-5/2}\ .
\ee
Putting everything together, we get
\be
\frac{1}{16\sqrt 2 |x|^{1/2}|1-x|^{4}} \sqrt{1+|x|+|1-x|} \ .
\ee

\item $G^-G^+$ in the first and $\bar G^- \bar G^+$ in the second factor:
The first factor is
\be
\langle N^1_{1;1}(\infty) G^+_{-1/2} N^2_{-2;-2}(1) G^-_{-1/2} N^2_{2;2}(x)N^1_{-1;-1}(0)\rangle\ ,
\ee
giving
\be
B_{-221}(x)B_{2-21}(\bar x)= x^{1/4}\bar x^{-1/4}|1-x|^{-1}
\ee
for the bosonic part and for the Ising part
\be
\frac12 F_{\psi\psi}(1,x)=\frac14\frac1{1-x}(x^{1/2}+x^{-1/2})\ .
\ee
The second factor is 
\be
\langle N^1_{-1;-1}(\infty) \bar G^+_{-1/2} N^1_{-1;-1}(1) \bar G^-_{-1/2}N^1_{1;1}(x)N^1_{1;1}(0)\rangle
\ee
giving a bosonic part
\be B_{1-1-1}(x)B_{-33-1}(\bar x)=x^{1/8}(1-x)^{-1/8} \bar x^{-3/8}(1-\bar x)^{-9/8}\ ,
\ee
and from the spin fields
\be
\frac14F_{\sigma\mu\mu\sigma}(x,\bar x)\ .
\ee
The third factor is simply
\be
|1-x|^{-1/2}\ .
\ee

Putting everything together, we get
\be
\frac{(x^{1/2}+x^{3/2})}{32 |x|^{3/2}|1-x|^{4}} \left((1-\sqrt{1-x})^{1/2}(1+\sqrt{1-\bar x})^{1/2} + (1+\sqrt{1-x})^{1/2}(1-\sqrt{1-\bar x})^{1/2}\ .
\right)
\ee

\item $G^-G^+$ in the second and $\bar G^- \bar G^+$ gives the complex conjugate of the correlator above.

\item $G^-G^+$ in the first and $\bar G^- \bar G^+$ in the third factor:
The first factor is again
\be
\langle N^1_{1;1}(\infty) G^+_{-1/2} N^2_{-2;-2}(1) G^-_{-1/2} N^2_{2;2}(x)N^1_{-1;-1}(0)\rangle\ ,
\ee
giving
\be
B_{-221}(x)B_{2-21}(\bar x)= x^{1/4}\bar x^{-1/4}|1-x|^{-1}
\ee
for the bosonic part and for the Ising part
\be
\frac12 F_{\psi\psi}(1,x)=\frac14\frac1{1-x}(x^{1/2}+x^{-1/2})\ .
\ee
The second factor is 
\be
\langle N^1_{-1;-1}(\infty) N^1_{-1;-1}(1) N^1_{1;1}(x)N^1_{1;1}(0)\rangle\ ,
\ee
giving a bosonic part
\be B_{1-1-1}(x)B_{1-1-1}(\bar x)
\ee
and from the spin fields
\be
F_{\sigma\sigma\sigma\sigma}(x,\bar x)\ .
\ee
The third factor is simply
\be
\frac14 (1-x)^{-1/4}(1-\bar x)^{-5/4}\ .
\ee
Putting everything together, we get
\be
\frac{(1+x)}{16\sqrt 2 |x|^{1/2}|1-x|^{4}} \sqrt{1+|x|+|1-x|} \ . 
\ee

\item $G^-G^+$ in the third and $\bar G^- \bar G^+$ in the first factor gives the same as above but with $x \leftrightarrow\bar x$.

\item $G^-G^+$ in the third and $\bar G^- \bar G^+$ in the second factor: 
The first factor is
\be
\langle N^1_{1;1}(\infty) N^2_{-2;-2}(1) N^2_{2;2}(x)N^1_{-1;-1}(0)\rangle\ ,
\ee
giving
\be
B_{2-21}(x)B_{2-21}(\bar x)
= |x|^{-1/2}|1-x|^{-1}\ .
\ee
The second factor is 
\be
\langle N^1_{-1;-1}(\infty) \bar G^+_{-1/2} N^1_{-1;-1}(1) \bar G^-_{-1/2}N^1_{1;1}(x)N^1_{1;1}(0)\rangle\ ,
\ee
giving a bosonic part
\be B_{1-1-1}(x)B_{-33-1}(\bar x)
\ee
and an Ising part
\be
F_{\sigma\mu\mu\sigma}(x,\bar x)/4\ .
\ee
The third factor is simply
\be
(1-x)^{-5/4}(1-\bar x)^{-1/4}/4\ .
\ee
Putting everything together, we get
\be
\frac{x^{1/2}}{32 |x|^{3/2}|1-x|^{4}} \left((1-\sqrt{1-x})^{1/2}(1+\sqrt{1-\bar x})^{1/2} + (1+\sqrt{1-x})^{1/2}(1-\sqrt{1-\bar x})^{1/2}
\right)\ .
\ee
\item $G^-G^+$ in the second and $\bar G^- \bar G^+$ in the third factor gives the complex conjugate of the above.

\end{enumerate}

\subsubsection{$O(1)O^\dagger(x)$}
From how hermitian conjugation works, we get the same correlation function as in the case $O^\dagger(1)O(x)$, except that the bosonic momenta of $O$ and $O^\dagger$ pick up a minus sign. That is, we can replace $B_{k_2k_3k_4}$ by $B_{-k_2-k_3k_4}$ in the above expressions.

\subsection{Integrating the overall contribution}
Let us now discuss how to integrate the overall result for the correlator 
\be
G(x)= 
\langle \varphi_{12}(\infty) \Phi(1)\Phi(x) \varphi_{12}^\dagger(0)\rangle\ ,
\ee
which is given by summing up all contributions above and then dividing by 2 due to the factor of $\sqrt{2}$ in the definition of $\Phi$.
We note that for $x\to \infty$ we have
\be
G(x) = \frac1{32} |x|^{-2} +\ldots\ = \frac{C_{\varphi_{31}^\dagger \Phi \varphi_{12}}^2 }{|x|^2} +\ldots\ ,
\ee
where the leading term comes from term 1 in \ref{sss:OdO}. This gives the expected logarithmic divergence discussed at the end of section~\ref{ss:2ndpert}.

We did not attempt to evaluate the resulting integral analytically, but instead used Mathematica to approximate it. For this, we first use the transformation $x\mapsto 1-x$ to move the non-integrable singularities to $x=0$. (In the one case where that leads to a singularity at $x=1$, we manually subtract a correction term $\frac1{32} |x|^{-2}$ to remove the singularity, and then add the integral of the correction term back in.) We then introduce polar coordinates and split the integral into two integrals, one from $r=\varepsilon$ to 1, the other from $r=1$ to $\varepsilon^{-1}$. In each integral we expand the correlator in a power series in $r$ and $r^{-1}$ respectively, and integrate term by term. For our results we kept terms up to the 15th power. Experimentally we found that the value of the integral has converged pretty well by that point, so we are fairly confident that including higher power terms will not change the answer significantly. A Mathematica notebook of this computation can be found in the supplements to this article.

\begin{figure}[h!]
	\centering\includegraphics[width=.6\textwidth]{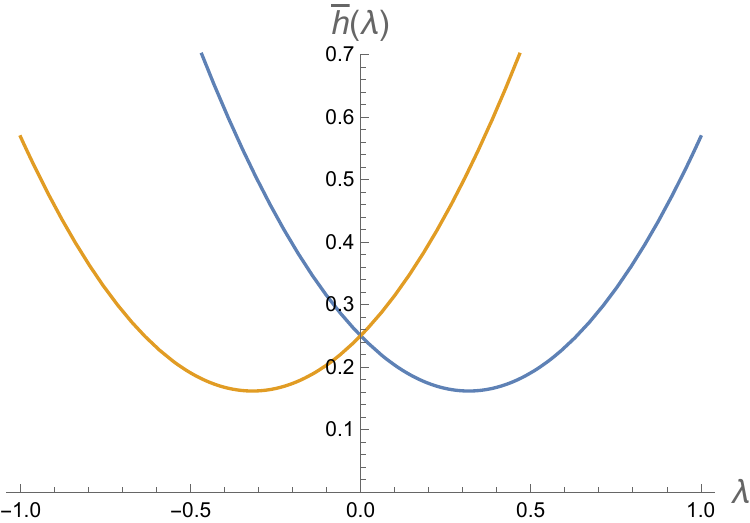}
	\caption{ $\bar h$ as a function of the coupling constant $\lambda$ for the two lifted fields.}
	\label{fig:plot}
\end{figure}

This procedure leads to the result 
\be
\int G(x) d^2x = \frac{\pi }{\varepsilon ^2}+\frac{\pi }{8\varepsilon }-  \frac{ \pi}{8}  \log (\varepsilon ) +M + \ldots\ ,
\ee
where
\be
M = -0.56\ldots\ .
\ee
As explained in section~\ref{ss:2ndpert},  $M$ then gives the second order lifting matrix for both $\varphi_+$ and $\varphi_-$. Their weights $\bar h_\pm(\lambda)$ are thus approximated at second order by
\be
h_\pm(\lambda)= \frac14 \mp \frac{\pi}{4\sqrt2}\lambda + \frac{\pi}2 0.56 \lambda^2 + O(\lambda^3)\ .
\ee
Taking this second order approximation at face value, the fields $\varphi_\pm$ have a minimal weight $\bar h=0.16$ at $\lambda=\pm0.32$. Comparing the contribution of the three terms in this quadratic approximation, each power suppresses the contribution by about 1/2, so our quadratic approximation may not be too far off from the full result.

\section*{Acknowledgements}
We thank Luis Apolo, Nathan Benjamin, Suzanne Bintanja and Ida Zadeh  for helpful discussions and comments on the draft. We thank an anonymous referee for useful comments on a previous version of the article.
The work of CAK is supported in part by NSF Grant 2111748.

\appendix
\section{Evaluating integrals} \label{app:int}
In this appendix we collect some methods for evaluating integrals of the form (\ref{Iint}) and (\ref{Jint}). Even though it turns out that for our values of $a$ and $b$ we can evaluate the integrals using elementary methods, the methods discussed here may be of future use.

As long as we are only interested in the constant part of the integral, equation (3.63) in \cite{MR972993} gives the value
\be
\int d^2 x |x|^{2a}|1-x|^{2b} = -\sin(\pi b) \frac{\Gamma(1+a)\Gamma(1+b)^2\Gamma(-1-a-b)}{\Gamma(2+a+b)\Gamma(-a)}\ ,
\ee
This answer is obtained by a clever complex deformation of the integration contours. Poles are effectively regularized by closing contours with segments at infinity and discarding the (divergent) contributions of these segments. Effectively this allows to deal with power divergences. However, for our values of $a,b$ the resulting $\Gamma$ functions are evaluated at poles and are therefore undefined. This is due to the appearance of logarithmic divergences in our integral. 

Another approach is to transform these integrals to polar coordinates and then first evaluate the angular integral. This leads to hypergeometric functions, which then in turn can be integrated over $r$ \cite{Apolo:2022pbq}.

The central formula here is that for $Re(c)>Re(b)>0$ 
\be\label{2F1int}
\int_0^{\pi/2} \frac{\sin(t)^{2b-1}\cos(t)^{2c-2b-1}}{(1-z\sin(t)^2)^{a}} dt
= \frac{\Gamma(b)\Gamma(c-b)}{2\Gamma(c)}{}_2F_1(a,b,c;z)\ .
\ee
To evaluate
\be
\int d^2 x |x|^{2a}|1-x|^{2b}\ ,
\ee
we assume that $b<-1$, so that any divergence at $x=1$ is integrable and therefore does not need to be regularized.
Switching to polar coordinates then gives
\begin{multline}
\int_0^{2\pi} d\theta r^{2a+1}|1+2r\cos\theta +r^2|^{b}
=  r^{2a+1}\int_0^{2\pi} d\theta |1+2r +r^2-4r \sin^2\frac\theta2|^{b}\\
= r^{2a+b+1}(\alpha+2)^{b}\int_0^{2\pi} d\theta |1-z \sin^2\frac\theta2|^{b}
= r^{2a+b+1}(\alpha+2)^{b}4\int_0^{\pi/2} dt |1-z \sin^2t|^{b} \ ,
\end{multline}
where $\alpha=r+1/r$ and $z=4/(\alpha+2)$. Using (\ref{2F1int}) we get
\be
r^{2a+b+1}(\alpha+2)^{b} 2 \frac{\Gamma(1/2)^2}{\Gamma(1)} {}_2F_1(-b,1/2,1;z)
=2\pi r^{2a+b+1}(\alpha+2)^{b} {}_2F_1(-b,1/2,1;\frac4{\alpha+2})\ .
\ee
Next we rewrite the integrand by using the fact that it is symmetric under $r\to 1/r$:
\be
\int_\varepsilon^1 dr 2\pi (r^{2a+b+1}+r^{-2a-b-3})(\alpha+2)^{b} {}_2F_1(-b,1/2,1;\frac4{\alpha+2})
\ee
To simplify the argument of the hypergeometric function, we then use Kummer's quadratic transformation (since $r<1$):
\be
{}_2F_1(-b,1/2,1;\frac{4r}{(1+r)^2})=(1+r)^{-2b}{}_2F_1(-b, -b,1; r^2)\ .
\ee
The $r$-integral is thus
\be
\int_\varepsilon^1 dr 2\pi r^{2a+1}(1+r^{-4a-2b-4} ){}_2F_1(-b,-b,1;r^2)\ .
\ee
This integral can then be evaluated either in closed form or term by term in $r$.

We can similarly evaluate the integral
\be
-\int d^2 x |x|^{2a}|1-x|^{2b}(x+\bar x)\ .
\ee
In polar coordinates we have
\begin{multline}
\int_0^{2\pi}d\theta 2r^{2a+2}|1+2r\cos\theta+r^2|^{b}\cos\theta
= 2r^{2a+b+2}(\alpha+2)^{b}\int_0^{2\pi} d\theta |1-z \sin^2\frac\theta2|^{b}(1-2\sin^2\frac\theta 2)\\
= r^{2a+b+2}(\alpha+2)^{b} 4 \left(\frac{\Gamma(1/2)^2}{\Gamma(1)} {}_2F_1(-b,1/2,1;z)-2\frac{\Gamma(3/2)\Gamma(1/2)}{\Gamma(1)} {}_2F_1(-b,3/2,2;z)\right)\\
= 4\pi r^{2a+b+2}(\alpha+2)^{b}  \left( {}_2F_1(-b,1/2,1;z)- {}_2F_1(-b,3/2,2;z)\right)\ .
\end{multline}
This can then be integrated over $r$, if necessary term by term.

However, as mentioned above, the specific types of integrals that we are interested in actually have a far simpler form. We have
\bea
I(-2,0) &=& \int d^2x |x|^{-4} = \int dr 2\pi r^{-3} = \pi(\epsilon^{-2}-\epsilon^2)\ ,\\
I(-2,1) &=& \int d^2x |x|^{-4} |1-x|^{2} = \int dr 2\pi (r^{-3}+r^{-1})
=\pi  ( \epsilon ^{-2}-\epsilon ^2-4 \log \epsilon)\ .
\eea
Finally the integrand of $J(-2,0)$ is simply $2r^{-3}\cos\theta$, so that after the $\theta$ integration we immediately have $J(-2,0)=0$.

\bibliographystyle{ytphys}
\bibliography{refmain}

\end{document}